\documentclass[twocolumn]{aastex631}

\let\splitbox\relax 
\usepackage{adjustbox}

\usepackage{xcolor}
\usepackage{subfloat}
\usepackage{multirow}
\usepackage{soul}

\usepackage{graphicx}	
\usepackage{amsmath}	

\usepackage{amssymb}	
\usepackage{cleveref}   
\usepackage{ulem}
\usepackage{multirow}

\shorttitle{Two Sub-Neptunes in the K2-266 Planetary System}
\shortauthors{Jiang et al.}

\begin{document}

\title{Towards High Precision Mass Measurements of Two Sub-Neptunes in the K2-266 Planetary System Through Transit Timing} 


\author[0000-0001-7359-3300]{Ing-Guey Jiang}
\thanks{This study uses CHEOPS data observed as the Guest Observers\\  (GO) programme CH\_PR230008 (PI Ing-Guey Jiang).}
\affiliation{Department of Physics and Institute of Astronomy, National Tsing-Hua University, Hsinchu 30013, Taiwan}
\email{jiang@phys.nthu.edu.tw}

\author[0000-0001-8677-0521]{Li-Chin Yeh}
\affiliation{Institute of Computational and Modeling Science, National Tsing-Hua University, Hsinchu 30013, Taiwan}

\author[0000-0002-5494-3237]{Billy Edwards}
\affiliation{SRON, Netherlands Institute for Space Research, Niels Bohrweg 4, NL-2333 CA, Leiden, The Netherlands}

\author[0000-0002-6926-2872]{Ming Yang}
\affiliation{
College of Surveying and Geo-Informatics, 
Tongji University, Shanghai, 
200092, People's Republic of China}

\author[0000-0002-3481-9052]{Keivan G. Stassun}
\affiliation{Department of Physics and Astronomy, Vanderbilt University, Nashville, TN 37235,  USA}

\author[0000-0001-7234-7167]{Napaporn A-thano}
\affiliation{National Astronomical Research Institute of Thailand, Chiang Mai, 50180, Thailand}

\begin{abstract}
Sub-Neptunes have been found to be one of the most common types of exoplanets, yet their physical parameters and properties
are poorly determined and in need of further investigation. 
In order to improve the mass measurement and parameter determination of two sub-Neptunes, 
K2-266~d and K2-266~e,
we present new transit observations obtained with CHaracterising ExOPlanets Satellite (CHEOPS) and 
Transiting Exoplanet Survey Satellite (TESS), 
increasing the baseline of transit data from a few epochs to 
165 epochs for K2-266~d, and to 121 epochs for K2-266~e.
Through a two-stage fitting process, 
it is found that the masses of K2-266~d and K2-266~e are 
6.01$\pm$0.43 $M_\oplus$ and  
7.70$\pm$0.58 $M_\oplus$, respectively. 
With these updated values and one order of magnitude better
precision, we confirm the planets to belong to the population of planets
that has been determined to be volatile-rich.
Finally, we present the results of dynamical simulations, showing that 
the system is stable, the orbits are not chaotic, and that
these two planets are close to but not in 4:3 mean motion resonance.
\end{abstract}


\section{Introduction}
\label{sec:intro}

The rapid advancement of exoplanet astronomy has significantly influenced the fields of planetary science and planet formation in numerous ways. 
The properties of detected exoplanets offer valuable insight into the potential
configurations of
all planet populations. Moreover, the presence of these exoplanets and the structures of extrasolar planetary systems offer clues regarding the processes of planet formation and evolution. Despite the intriguing and exciting ongoing progress, much remains to be learned in order to obtain a comprehensive understanding of how these objects form and evolve.

To advance towards the goal of constructing a cohesive and complete understanding, delving deeper into characterizing planetary systems that represent extremes of parameter space, using the current available data and techniques, holds promise. By gaining a comprehensive understanding of these distinctive systems, we can then proceed to explore the mechanisms underlying their formation, thereby enriching our theoretical framework of planet formation.

A large fraction of known transiting exoplanets were detected by the Kepler space telescope\citep{Koch2010},
now numbering
several thousands\citep{Borucki+2010Sci}. 
Upon the failure of Kepler's second reaction wheel,
K2 campaigns were organized to use 
the Kepler Space Telescope to observe a set of 
fields along the ecliptic \citep{Howell2014PASP}, further increasing the number and diversity of Kepler exoplanet discoveries. 
Among these exoplanets, it was 
statistically determined that 
sub-Neptunes and super-Earths 
represent the most common outcome of the planet formation and evolution process\citep{Latham2011, Dorn2015}.
The definition of sub-Neptunes and super-Earths is that 
planets have masses between one Earth mass and 
17.15 Earth mass (Neptune mass). However, the boundary between
super-Earths and sub-Neptunes is complicated and depends on other planetary characteristics. 
According to \citet{Parc2024}, the maximum mass of super-Earths is close to 10 $M_\oplus$, 
but the minimum mass of sub-Neptunes could be 
between 1.9 $M_\oplus$ and 4.3 $M_\oplus$.
Therefore, further investigations on the properties of these sub-Neptunes
\citep{Valencia2013, Wang2014, Lopez2014, Mortier2016, Lammer2016, Kubysh2019, Kite2019, Benneke2019, Nixon2021, Luque2023, Palethorpe2024} and super-Earths
\citep{Thomas2016, Raymond2018, 
Hirano2021, Morris2021, Moore2024, Gajendran2024} 
would be helpful in developing
a complete picture of planet formation.

The star K2-266 was observed during K2 Campaign 14 in 2017. 
The K2-266 planetary system is a compact, misaligned multi-planet system discovered by \citet{Rodri2018AJ}.
Among the six planets presented 
by those authors,
the planet K2-266~d and the planet K2-266~e were confirmed 
through transit timing variations (TTVs), 
the planets K2-266~b and K2-266~c 
are validated, and another two 
(K2-266.02, K2-266.06) are candidates. 
The main planetary parameters of all the above six exoplanets
are summarized in Table~\ref{table:main}.
The planet K2-266~b has a very short orbital period, i.e.,
around 0.66 days, so it is an ultrashort-period (USP) planet. 
USP is defined as planets with orbital periods of less than 1 day \citep{Sanchis2014}. 
This USP planet has an inclination angle around 75 degrees and
all other planets have inclinations around 88 or 89 degrees, so it is a misaligned system with approximately two inclined 
orbital planes. 

\begin{table*}
\caption{
The main planetary parameter of all six planets adopted from Table 4 in \citet{Rodri2018AJ}.}
\label{table:main}
\hspace{-2.4cm}
\begin{adjustbox}{width=1.13\textwidth}
\begin{tabular}{|c|c|c|c|c|c|c|}\hline
Name & K2-266 b & K2-266.02 & K2-266 c & K2-266 d & K2-266 e & K2-266.06 \\ \hline
orbital period (day) & $0.658524\pm 0.000017$ & $6.1002_{-0.0017}^{+0.0015}$ & $7.8140_{-0.0016}^{+0.0019}$&  $14.69700_{-0.00035}^{+0.00034}$  & $19.4820\pm 0.0012$ & $56.682_{-0.018}^{+0.019}$ \\ \hline
semi-major axis (AU)& $0.01306_{-0.00021}
^{+0.00020}$ & $0.05761_{-0.00093}^{+0.00090}$   & $0.0679\pm 0.0011$ & $0.1035_{-0.0017}^{+0.0016}$ & $0.1249_{-0.0020}^{+0.0019}$ & $0.2546_{-0.0041}^
{+0.0040}$  \\ \hline
eccentricity   & ... & $0.051_{-0.036}^{+0.051}$ & $0.042_{-0.030}^{+0.043}$ & $0.047_{-0.032}^{+0.043}$ & $0.043_{-0.030}^{+0.036}$ & 
$0.31_{-0.17}^{+0.11}$\\ \hline
argument of periastron (degree) & ... &  $88_{-62}^{+60}$ & $87\pm 61$ &$87\pm 62$ & $89_{-58}^{+57}$ &  $83_{-59}^{+57}$ \\ \hline
inclination (degree) & $75.32_{-0.70}^{+0.62}$ & 
 $87.84_{-0.46}^{+0.84}$ & $88.28_{-0.41}^{+0.81}$ 
 &$89.46_{-0.25}^{+0.32}$
 &$89.45_{-0.18}^{+0.25}$ 
 &$89.40_{-0.14}^{+0.26}$ \\ \hline
mass ($M_\oplus$) & $11.3_{-6.5}^{+11}$ 
& $0.209_{-0.089}^{+0.15}$ 
& $0.29_{-0.11}^{+0.17}$ 
& $9.4_{-2.0}^{+2.9}$ 
& $8.3_{-1.8}^{+2.7}$ 
& $0.70_{-0.30}^{+0.87}$ \\ \hline
radius ($R_\oplus$) & $3.3_{-1.3}^{+1.8}$ 
& $0.646_{-0.091}^{+0.099}$  
& $0.705_{-0.085}^{+0.096}$  
& $2.93_{-0.12}^{+0.14}$  
& $2.73_{-0.11}^{+0.14}$ 
& $0.90_{-0.12}^{+0.14}$  \\ \hline
\end{tabular}
\end{adjustbox}
\end{table*}

\begin{table*}
\caption{The Log of CHEOPS Observations.
The exposure time is 60 sec for all frames in all visits. Visit d1, d2, d3 are for the planet K2-266~d, and Visit e1, e2, e3 are for the planet K2-266~e. 
The visit ID, the starting time of observations, the ending time of observations, the duration of observations, 
the number of frames, and the file key 
are listed in different columns successively.
The data can be retrieved from the CHEOPS archive by 
the target name.}
\label{LogCHEOPS}
\hspace{-0.25cm}
\begin{adjustbox}{width=0.9\textwidth}
\begin{tabular}{|c|c|c|c|c|c|}\hline
Visit & Start Date & End Date & Duration & Number of & File Key\\ 
& [UTC] & [UTC] & [hours] & Frames &    \\ \hline 
d1 & 2023-02-26 19:52 & 2023-02-27 07:18 
& 11.44 & 482 & CH\_PR230008\_TG000101\_V0300\\ \hline
d2 & 2023-04-26 15:02 &2023-04-27 01:52
& 10.84 & 359 & CH\_PR230008\_TG000102\_V0300\\ \hline
d3 & 2024-01-30 22:04 &2024-01-31 10:33 
& 12.49 & 462 & CH\_PR230008\_TG000103\_V0300\\ \hline
e1 & 2023-03-19 12:03 & 2023-03-19 23:26
& 11.39 & 630 & CH\_PR230008\_TG000201\_V0300\\ \hline
e2 & 2023-04-07 22:54 & 2023-04-08 10:20
& 11.44 & 483 & CH\_PR230008\_TG000202\_V0300\\ \hline
e3 & 2023-04-27 10:49 & 2023-04-27 21:38
& 10.82 & 368 & CH\_PR230008\_TG000203\_V0300\\ \hline
\end{tabular}
\end{adjustbox}
\end{table*}

\begin{table*}
\caption{
The stellar parameters adopted from \citet{Rodri2018AJ}.}
\label{table:origin-value}
\hspace{0.8cm}
\begin{adjustbox}{width=0.8\textwidth}
\begin{tabular}{|c|c|c|c|}\hline
Star & surface gravity ${\rm log} g$ (cgs) & effective temperature (K)  & metallicity [Fe/H]  \\ \hline
K2-266  & $4.581_{-0.037}^{+0.032}$ & $4285_{-57}^{+49}$ 
& $-0.12_{-0.42}^{+0.40}$ \\ \hline
\end{tabular}
\end{adjustbox}
\end{table*}

\begin{table*}
\caption{The log of TESS observations. The exposure time is 120 sec for all frames. The sector numbers, the starting time of observations, the ending time of observations, the number of frames, the transit epoch ID of the planet K2-266~d, and the transit epoch ID of the planet K2-266~e are listed in different columns successively.}
\label{LogTESS}
\hspace{-0.5cm}
\begin{adjustbox}{width=0.9\textwidth}
\begin{tabular}{|c|c|c|c|c|c|}\hline
Sector  & Start Date & End Date &  Number of PDCSAP & Epoch of   & Epoch of \\
        & [UTC]  & [UTC] & Data Points  & K2-266 d & K2-266 e \\ \hline
    35  &   2021-02-09 22:06 &  2021-03-06 11:37 &    13689  & none & 69\\ \hline   
   45   &   2021-11-07 00:05& 2021-12-02 02:59 &  15682   &110,111 & 83 \\ \hline
   46   &   2021-12-03 01:33 & 2021-12-30 04:52 &  16425  &112,113 & 84,85 \\ \hline
   62   &   2023-02-12 22:40& 2023-03-10 15:46  &  16005  & none & 107\\ \hline
   72 & 2023-11-11 16:19 & 2023-12-07 01:45 & 13693 & 160,161 & 121
   \\ \hline
\end{tabular}
\end{adjustbox}
\end{table*}

\begin{figure*}
\centering
\includegraphics[width=0.9\textwidth]{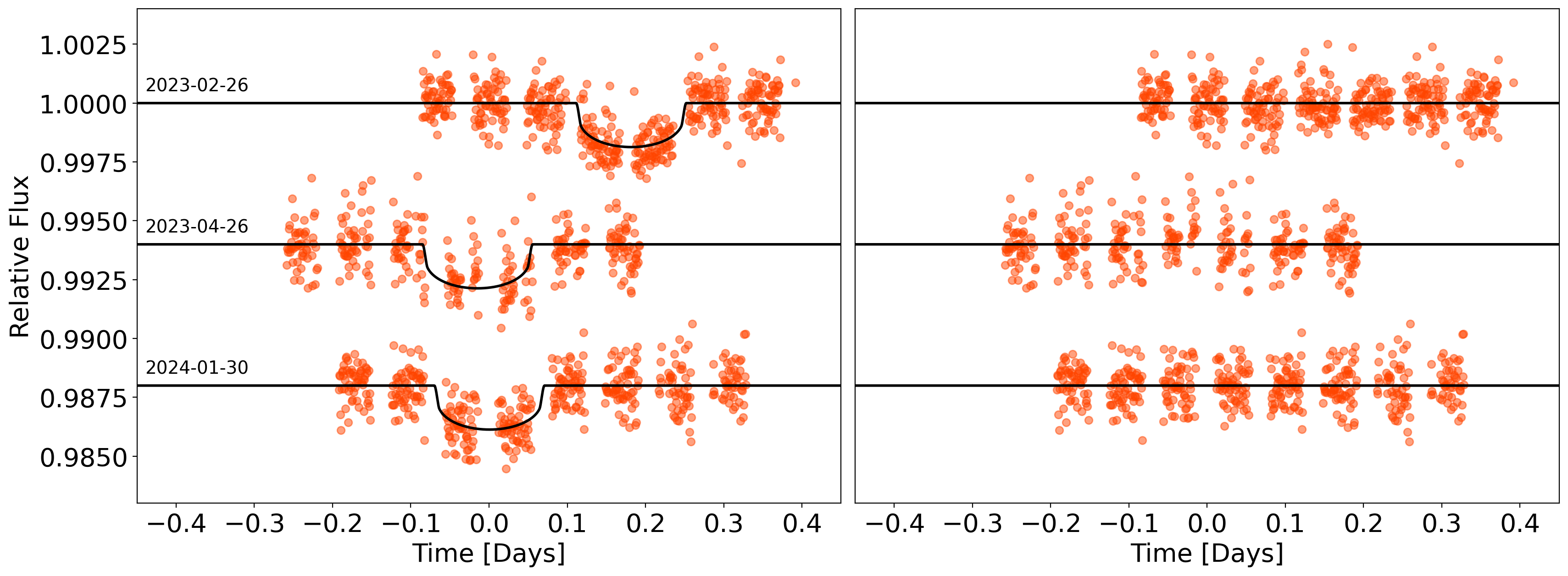}
\includegraphics[width=0.9\textwidth]{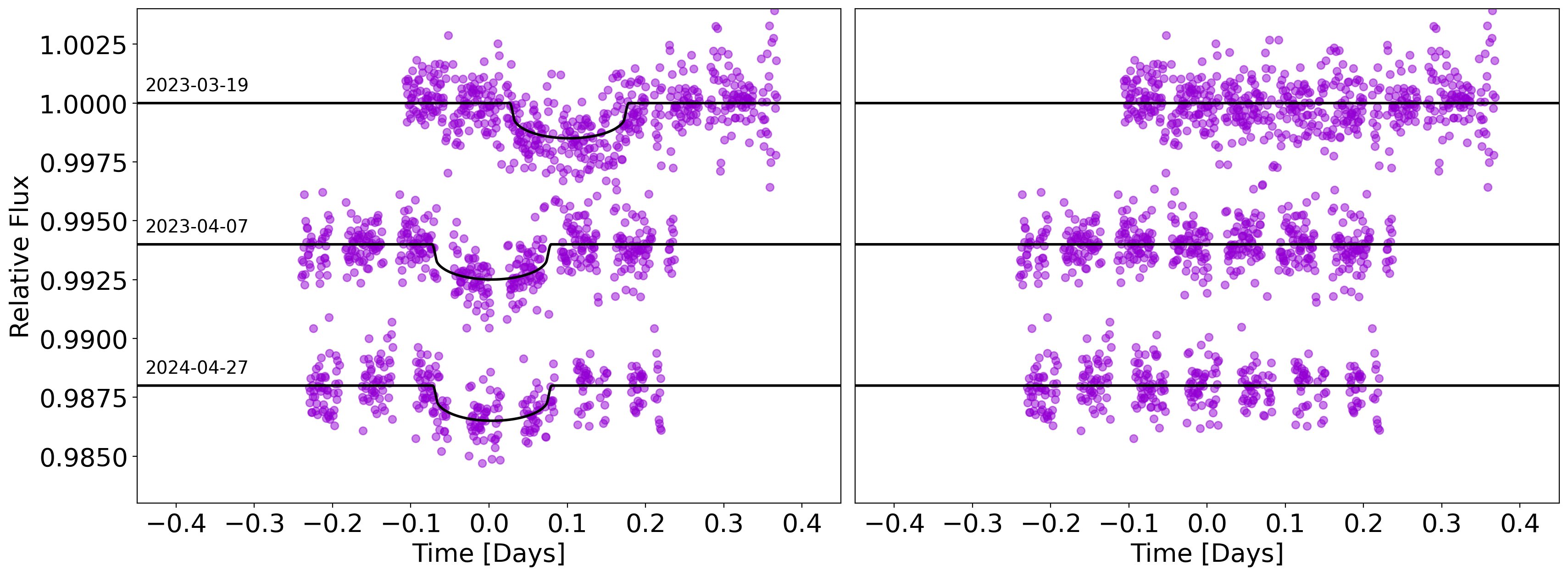}
\caption{The CHEOPS transit light curves of planet K2-266 d (top, orange) and K2-266 e (bottom, purple). For each planet, we show the corrected light curve and best-fit model (left) as well as the residuals (right). The data are plotted with respect to the linear ephemeris from \citet{Rodri2018AJ}.}
\label{figCHEOPSLC}
\end{figure*}

\begin{figure*}
\centering
\includegraphics[width=0.85\textwidth]{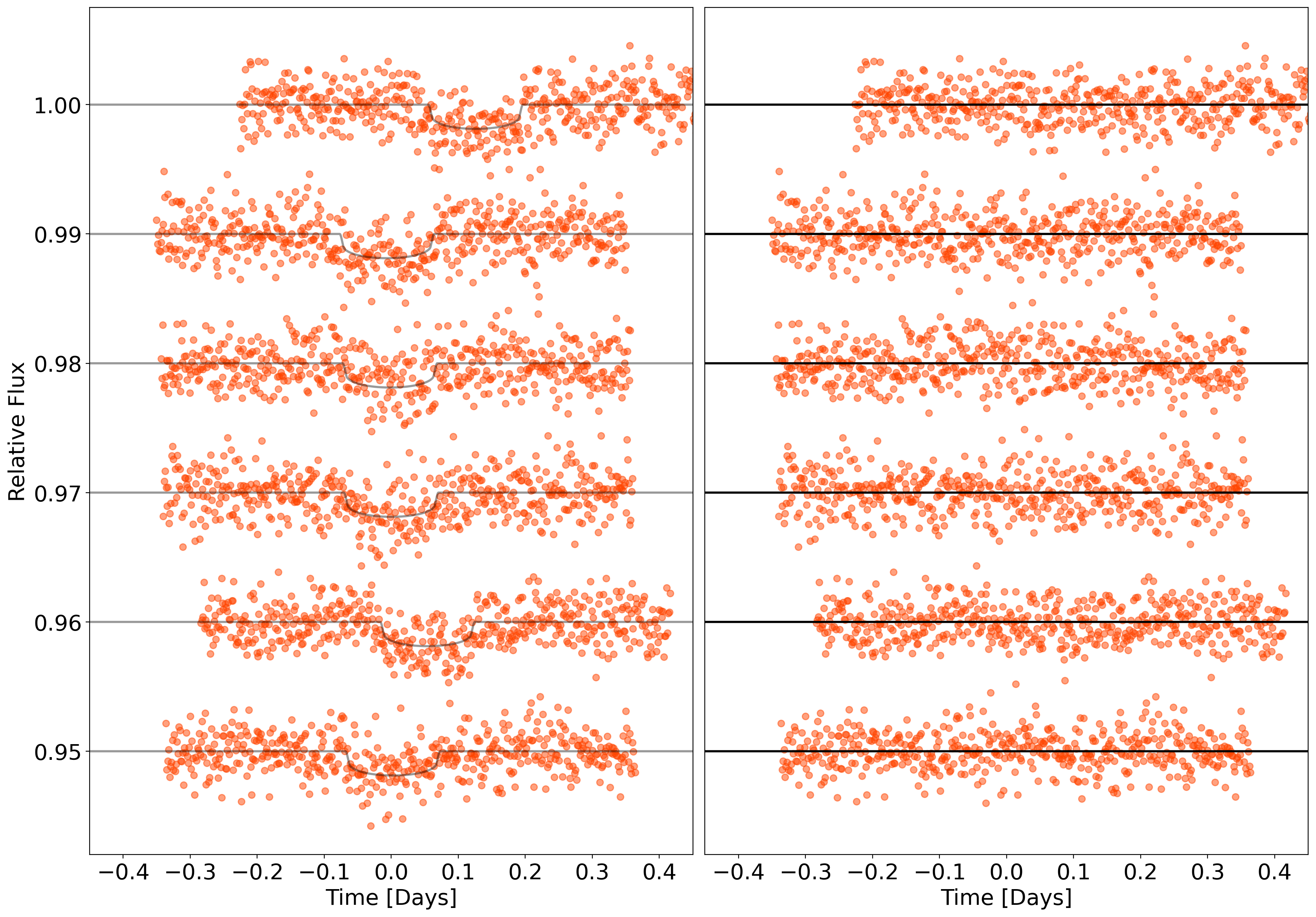}
\includegraphics[width=0.85\textwidth]{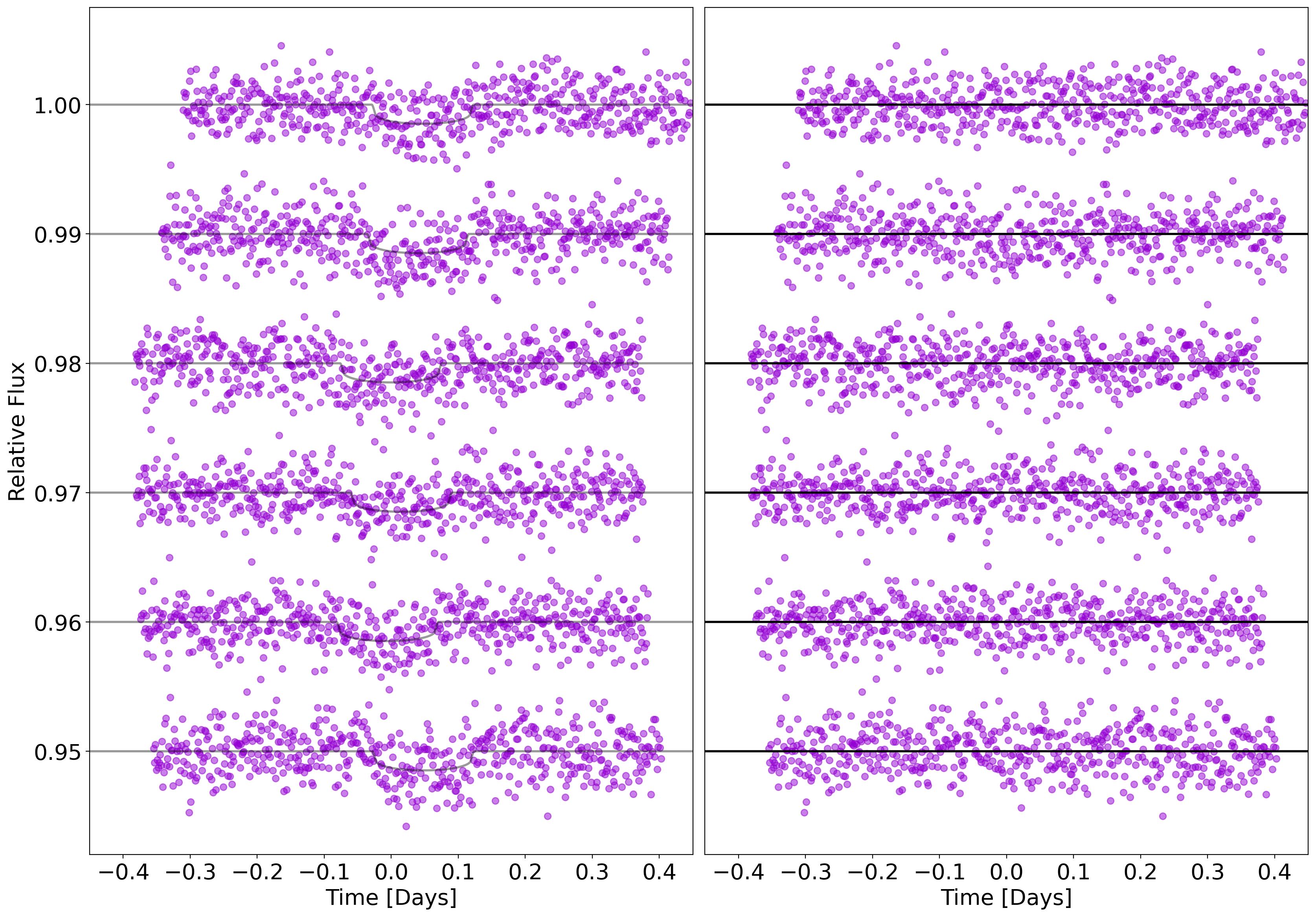}
\caption{The TESS transit light curves of planet K2-266~d (top, orange) and K2-266~e (bottom, purple). For each planet, we show the corrected light curve and best-fit model (left) as well as the residuals (right). The data are plotted with respect to the linear ephemeris from \citet{Rodri2018AJ}.}
\label{figTESSLC}
\end{figure*}

\begin{table*}
\caption{The mid-transit times of the planet K2-266~d. The epoch IDs, mid-transit times, data sources, and references are listed in different columns successively.}
\label{tab:Planetd}
\hspace{2.05cm}
\begin{adjustbox}{width=0.66\textwidth}
\begin{tabular}{|c|c|c|c|}\hline
Epoch ID&  T ($BJD_{TDB}$) & Data Source & Reference\\ \hline
0   &  $2457915.44761\pm 0.00106$     & K2 &  \cite{Rodri2018AJ}     \\ \hline
1        & $2457930.13813\pm 0.00101$ & K2 &   \cite{Rodri2018AJ}    \\ \hline
2        & $2457944.83597\pm 0.00137$  & K2 &  \cite{Rodri2018AJ}    \\ \hline
3        &  $2457959.52590\pm 0.00090$ & K2 &  \cite{Rodri2018AJ}    \\ \hline
4        & $2457974.23919\pm 0.00114$  & K2 &  \cite{Rodri2018AJ}    \\ \hline
110    &  $2459532.24121\pm {0.00209}$ & TESS & this work\\ \hline
111   &  $2459546.93183 \pm  0.00425$  & TESS & this work\\ \hline
112 &  $ 2459561.62697  \pm 0.00677  $  & TESS & this work\\ \hline
113 & $ 2459576.32390 \pm 0.00560  $   & TESS & this work\\ \hline
142  & $2460002.60103  \pm 0.00318 $   & CHEOPS & this work\\ \hline
146 & $2460061.37436  \pm 0.00155  $    & CHEOPS & this work\\ \hline 
160 &$2460267.13700   \pm  0.00229 $& TESS & this work\\ \hline
161 & $2460281.83635   \pm 0.00252 $ & TESS & this work\\ \hline
165 & $2460340.61785  \pm  0.00546 $   & CHEOPS & this work\\ \hline
\end{tabular}
\end{adjustbox}
\end{table*}

\begin{table*} 
\caption{The mid-transit times of the planet K2-266~e. The epoch IDs, mid-transit times, data sources, and references are listed in different columns successively.
}
\label{tab:Planete}
\hspace{2.05cm}
\begin{adjustbox}{width=0.66\textwidth}
\begin{tabular}{|c|c|c|c|}\hline
Epoch ID&  T ($BJD_{TDB}$) & Data Source & Reference\\ \hline
0   &  $2457919.05628\pm 0.00088$ &  K2 & \cite{Rodri2018AJ}       \\ \hline
1   & $2457938.54211\pm 0.00108$ &   K2 & \cite{Rodri2018AJ}        \\ \hline
2    & $2457958.03343\pm 0.00112$  & K2 &  \cite{Rodri2018AJ}       \\ \hline
3     & $2457977.50614\pm 0.00120$  & K2 & \cite{Rodri2018AJ}        \\ \hline
69  & $2459263.36386 \pm 0.00509$  & TESS &  this work   \\ \hline
83  &  $ 2459536.15176  \pm 0.00438$  & TESS & this work \\ \hline
84   & $2459555.63343 \pm  0.00776$  & TESS & this work \\ \hline
85 &  $2459575.13036 \pm  0.00618$  & TESS & this work \\ \hline
107 & $2460003.72884  \pm 0.00502$    & TESS & this work\\ \hline
108  & $2460023.21754  \pm 0.00272$   &  CHEOPS & this work \\ \hline
109 & $2460042.70279  \pm  0.00321$    & CHEOPS & this work  \\ \hline
110 & $2460062.18893 \pm  0.00330  $    & CHEOPS & this work\\ \hline
121 & $2460276.52460  \pm 0.00381 $ & TESS & this work\\ \hline
\end{tabular}
\end{adjustbox}
\end{table*}

This misaligned compact configuration provides a relatively unique opportunity \citet{Becker2020AJ}
to study the possible formation processes. 
\citet{Becker2020AJ} proposed two methods,
i.e., the effect of stellar oblateness and 
the effect of an additional unseen planet,
to explain the misalignment of the K2-266~b USP planet.
In addition, through dynamical 
simulations, \citet{Rodri2018AJ} also
found that the majority of possible orbits are chaotic, and a small fraction of these orbits 
shows that the planet K2-266~d, e 
are in 
mean motion resonance. To pin down the explanation for the misalignment and the dynamical 
properties of this planetary system in general,
the planetary masses are the key parameters.
Unfortunately, 
there were limitations to the constraints that could be placed over such a short baseline available to the discovery analyses, with the masses being measured as $8.9^{+5.7}_{-3.8} M_\oplus$  and 
$14.3^{+6.4}_{-4.0} M\oplus$ for the planet K2-266~d and the planet K2-266~e, respectively. 

Moreover, being a close analog of our Earth, the internal structures, atmospheres, temperatures, and habitability of sub-Neptunes and super-Earths have attracted a lot of attention. 
It is clear that these physical properties all depend on the masses and orbits of planets.
In particular, planetary masses almost completely 
determine their internal structures and atmospheres.
As reported in \cite{Boujibar2020},
the planetary mass can affect the internal density profiles as well as the
coexistence of a solid and a liquid core,
which can contribute to the maintenance of a magnetic field.
The out-gassing 
of super-Earths is also 
controlled by their masses \citep{Levi2014}. 
To understand the current configuration and the dynamical stability of such a compact system,
a higher precision measurement on mass and orbital parameters are necessary. 

On the other hand, in a global picture, 
the mass-radius relations \citep{Otegi2020MR, Edmondson2023, Parc2024},
will be further improved and understood if
there are more sub-Neptunes, such as K2-266~d and K2-266~e, whose
mass measurements could be significantly improved in precision. In fact, due to the
mass uncertainties of K2-266~d and e, they were unable to be included in the 
exoplanet catalog defined by \citet{Parc2024}.
This exoplanet catalog called {\it PlanetS catalog} is
a list of exoplanets with higher precision of 
mass and radius values and can be used to 
update the mass-radius relations. 
Only planets with relative measurement
uncertainties smaller than 25\% in mass were considered in {\it PlanetS catalog} \citep{Parc2024}.
In order to obtain more transit data and have a longer TTV baseline,
we proposed and later were awarded to use CHaracterising ExOPlanets Satellite (CHEOPS) to do
transit observations of K2-266~d and K2-266~e 
(Program Number: AO3-08, 
Primary Investigator: Ing-Guey Jiang). 
These observations were performed smoothly,
and the results will be presented in this paper.

The analysis of light curves and the determination of transit timings are 
described in 
Section \ref{sec:obser}. 
The three-stage process employed to search for best-fit solutions and
the determination of masses and orbital parameters
through TTV fitting would be in 
Section \ref{sec:bestfit}. 
Our two planets'
positions in the mass-radius plane among those planets of PlanetS Catalog would be shown and discussed in 
Section \ref{sec:mr}.
In Section \ref{sec:dynamics}, the results of dynamical simulations would be presented. Finally, the conclusion is provided in 
Section \ref{sec:conclusion}.

\section{Observations and Data Analysis}
\label{sec:obser}

In this work, we use photometric light curves to measure the transit timings of K2-266~d 
and K2-266~e. Several facilities have studied this system. We take the transit mid times from K2 from \citet{Rodri2018AJ}, but conduct fits to the new CHEOPS (Section \ref{sec:cheops}) and TESS (Section \ref{sec:tess}) data.

\subsection{CHEOPS}
\label{sec:cheops}
The CHEOPS space telescope, which was successfully launched in December 2019, is an ESA (European Space Agency) small-class mission \citep{Benz2021} in a low-Earth orbit, with an orbital period about 98.7 minutes.  
CHEOPS is dedicated to observing bright stars that are already known to host planets.  The goal is to observe the transit events of exoplanets and thus characterize the configurations and properties of exoplanets. It observes one system at a time to perform a precise characterization. CHEOPS data has been widely employed and many fruitful results were produced \citep{Deline2022, Smith2022, Wilson2022, Nasci2023, Borsato2024, Egger2024, Vivien2024, Rosario2024, Fridlund2024, Pagano2024, Singh2024}

Six observational sequences of the K2-266 system were taken by CHEOPS. Half of these were transit observations of planet d, while the other half were taken while planet e was transiting the host star. The data were taken at a cadence of 60 seconds and the observation log is given in Table~\ref{LogCHEOPS}. The raw data of each visit were automatically processed by CHEOPS Data Reduction Pipeline 
\citep[\textit{DRP}~v14.1.3, ][]{Hoyer2020}, which calibrates the data (bias, gain, flat-fielding) before performing several corrections (cosmic ray hits, background, and smearing correction). Subsequently, the photometric signal of the target is extracted. The \textit{DRP} extracts light curves based on various aperture sizes. Here, we selected the R20 aperture as it provided the light curves with the smallest root-mean-square of data scattering.

For each planet, we fitted the three visits simultaneously. While the planet-to-star radius ratio was shared across all three visits, the mid time of each transit was fitted independently. All other system parameters were taken as Gaussian priors. The Gaussian means
were adopted from the values in
\cite{Rodri2018AJ}, and the
Gaussian widths were set as the lengths of 
error bars 
in \citep{Rodri2018AJ}. These are given in 
Table \ref{table:main} and
Table \ref{table:origin-value}. 
The transit light curve was modeled using 
{\it PyLightcurve}
\footnote{\url{https://github.com/ucl-exoplanets/pylightcurve}}   
\citep{Tsiaras2016ApJ}. 
For the limb darkening coefficients, Claret coefficients \citep{claret} were used and these were calculated using ExoTETHyS \citep{morello_exotethys}. 
In addition to these system parameters, we also fit systematics models for each dataset. CHEOPS has complex systematic trends, with the dominant trends being correlated with time and with the roll angle ($\Phi$) of the spacecraft \citep[e.g.,][]{maxted_pycheops}. As with previous studies, we apply linear decorrelation parameters with time and the background flux. To account for the systematics correlated with the roll-angle, we included a common cubic spline with breakpoints every $\sim$8 degrees, a technique which has previously been used for CHEOPS data \citep[e.g.,][]{osborn_hip9618}. These systematic corrections were applied separately for each visit.
The resulting CHEOPS light curves are presented
in Fig. \ref{figCHEOPSLC}.

\begin{figure*}[!htb]
\centering
\includegraphics[width=0.7\linewidth]{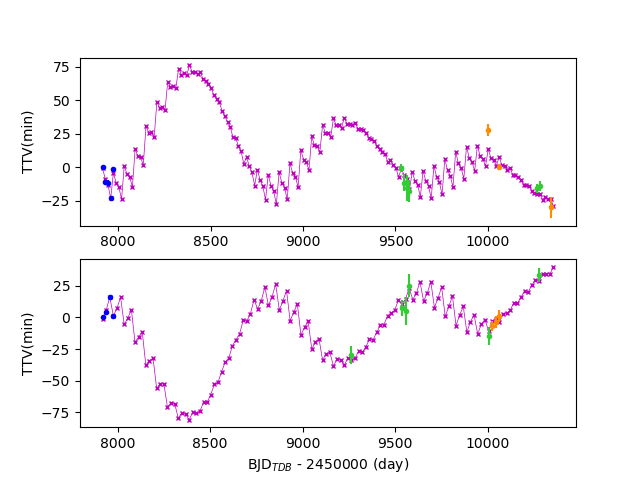}
\caption{TTV plots for the planet K2-266~d (top panel) and the planet K2-266~e (bottom panel). The points with error bars are observations, where blue ones are for K2, green ones are for TESS, and orange ones are for CHEOPS. The purple crosses and lines are the best-fit model (described in the later part of Section \ref{sec:bestfit}), in which the 1st cross for the planet K2-266~d is at the time 2457915.44712 $BJD_{TDB}$ and the 1st cross for the planet K2-266~e is at the time 2457919.05692 $BJD_{TDB}$.}
\label{fig:OC_org}
\end{figure*} 

\begin{table*}
\caption{
The Nelder-Mead fitting process}
\label{table:NMfitting}
\hspace{1.3cm}
\begin{adjustbox}{width=0.75\textwidth}
\begin{tabular}{|c|c|c|c|c|c|}\hline
 Parameter of K2-266 d& $m_d$ ($M_\oplus$)& $p_d$ (day)& 
$e_d$ & $\omega_d$ (degree)&  $Ma_d$ (degree)\\ \hline
Interval of Initial Value & (7.4, 12.3) &
(14.69665, 14.69734) & (0.015, 0.09) & (25, 149)& (0,360)  \\ \hline
Prior Distribution & uniform & uniform & 
uniform & uniform & uniform \\ \hline
Boundary & no & no & no & no & no \\ \hline\hline
Parameter of K2-266 e & $m_e$ ($M_\oplus$)& $p_e$ (day)& 
$e_e$ & $\omega_e$ (degree) &  $Ma_e$ (degree)\\ \hline
Interval of Initial Value& (6.5, 11) & 
(19.4808, 19.4832) & (0.013, 0.079) & 
(31, 146)  & (0,360)  \\ \hline 
Prior Distribution & uniform & uniform & uniform
 & uniform &  uniform\\ \hline
Boundary & no & no & no  & no  & no  \\ \hline
\end{tabular}
\end{adjustbox}
\end{table*}
\begin{table*}
\caption{
The MCMC fitting process}
\label{table:MCMCfitting}
\hspace{1.3cm}
\begin{adjustbox}{width=0.75\textwidth}
\begin{tabular}{|c|c|c|c|c|c|}\hline
Parameter of K2-266 d & $m_d$ ($M_\oplus$) & $p_d$ (day) & 
$ec_d$ & $es_d$ &  $Ma_d$ (degree)\\ \hline
Initial Value & 6.37 & 14.6913 & 
0.013 & 0.047&  349.84  \\ \hline
Prior Distribution & uniform & uniform & 
uniform & uniform & uniform \\ \hline
Boundary & no & no & no & no & no \\ \hline\hline
Parameter of K2-266 e & $m_e$ ($M_\oplus$) & $p_e$ (day) & 
$ec_e$ & $es_e$ & $Ma_e$ (degree) \\ \hline
Initial Value& 7.44  & 19.4943 & 0.00036 & 0.039  & 275.39  \\ \hline 
Prior Distribution & uniform & uniform & uniform
 & uniform &  uniform\\ \hline
Boundary & no & no & no  & no  & no  \\ \hline
\end{tabular}
\end{adjustbox}
\end{table*}

\begin{figure*}[!htb]
\centering
\includegraphics[width=1.0\linewidth]{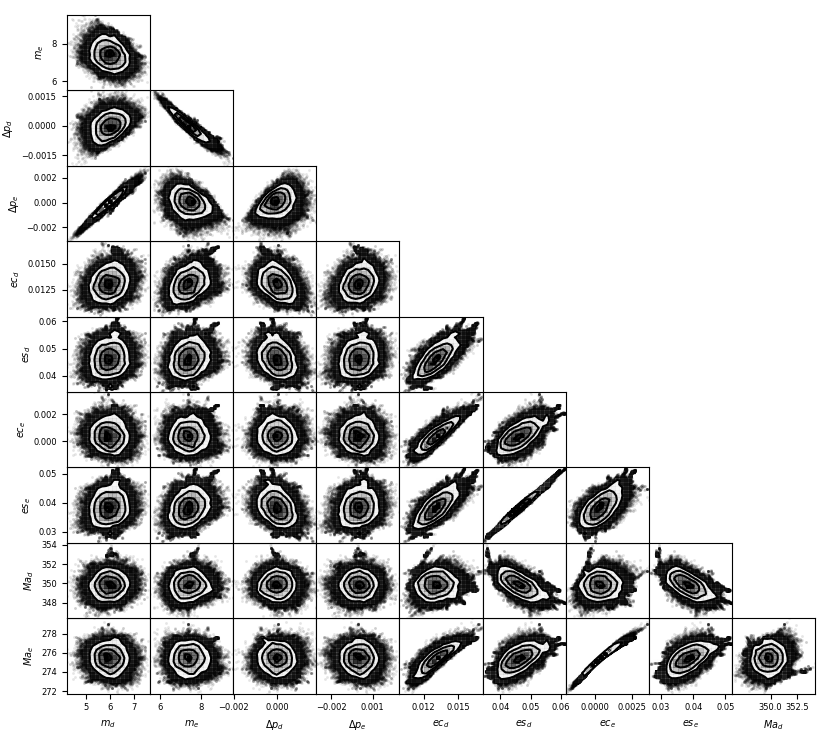}
\caption{The corner plot of 10-parameter MCMC fitting. For plotting convenience, we define $\Delta p_d\equiv p_d-14.69131$ and $\Delta p_e\equiv p_e-19.49367$ here.}
\label{fig:fit10_corner}
\end{figure*} 

\begin{table*}
\caption{The best-fit solution of 10-parameter MCMC fitting. The values of planetary mass $m$, orbital period $p$, $ec$, $es$,  mean anomaly $Ma$,
and corresponding orbital eccentricity $e$, 
the argument of pericenter $\omega$
for both planet K2-266~d and K2-266~e are listed.}
\label{table:bestMCMC10}
\hspace{-3.0cm}
\begin{adjustbox}{width=1.15\textwidth}
\begin{tabular}{|c|c|c|c|c|c|c|c|}\hline
Planet &$m(M_\earth)$&  $p$ (days)& $ec$ & $es$ & $Ma$(degrees) & $e$ & $\omega$ (degrees)\\ \hline
K2-266 d & 6.01 $\pm$ 0.43& 14.6911 $\pm$ 0.0005 &0.013 $\pm$ 0.001& 0.043 $\pm$ 0.006 &350.96 $\pm$ 1.43
& 0.045 $\pm$ 0.006& 73.10 $\pm$ 0.06\\ \hline
K2-266 e & 7.70 $\pm$ 0.58 & 19.4937 $\pm$ 0.0007 & 
0.00018  $\pm$ 0.00079 & 0.036 $\pm$  0.005 & 275.17 $\pm$ 1.02  & 0.036 $\pm$ 0.005
& 89.71 $\pm$ 0.02\\ \hline  
\end{tabular}
\end{adjustbox}
\end{table*}

\subsection{TESS}
\label{sec:tess}

In addition to the CHEOPS data, we also collect the transit light curves obtained by the Transiting Exoplanet Survey Satellite (TESS) for both planets. The star K2-266 was observed by TESS in Sectors 35, 45, 46, 62 and 72. We used the Pre-Search Data Conditioning Simple Aperture Photometry (PDCSAP) flux \citep{smith2017,smith_pdc2017}, which were calibrated through the Science Processing Operation Center (SPOC) pipeline 
\citep{Jenkins2016}.  These 2 minute light curves were downloaded from the Mikulski Archive for Space Telescopes 
(MAST)\footnote{\url{https://archive.stsci.edu/}}. We only used data points which we expected to be within a 0.4-day window for the planet K2-266~d and a 0.6-day window for the planet K2-266~e. These windows were centered at the expected mid-transit point using the linear ephemeris from \citet{Rodri2018AJ}. We extracted six transits for both planet K2-266~d and planet K2-266~e.
These were fitted in the same manner as the CHEOPS light curves, except for the systematic detrending for which we simply used a linear trend with time.
The TESS transit light curves are presented in Fig. \ref{figTESSLC}.
The TESS observation log is given in Table~\ref{LogTESS}, and we note that TESS will re-observe K2-266 in Sector 89.

The resulting mid-transit times of the planet K2-266~d
are presented in Table~\ref{tab:Planetd} and 
the mid-transit times of the planet K2-266~e
are listed in Table~\ref{tab:Planete}.
Note that the previously published values and corresponding uncertainties of mid-transit times,
which were observed by
K2 campaigns, are directly
adopted from \citet{Rodri2018AJ}. 
The standard unit of mid-transit time is 
the Barycentric Julian Date in the Barycentric Dynamical Time standard, i.e. $BJD_{TDB}$ \citep{Eastman2010}. 

\section{The Two-Stage Fitting Process on Transit Timings}
\label{sec:bestfit}

After obtaining the observational results of
mid-transit times for both 
planet K2-266~d and K2-266~e, the next step is to determine best-fit parameters through the 
dynamical fitting,
which could be a very complicated and difficult task. 
As in \cite{Hadden2017} and \cite{Rodri2018AJ},
the Markov Chain Monte Carlo (MCMC) sampling 
through the package {\it emecc}
\citep{Foreman2013} 
is a standard approach when dynamically determined 
theoretical transit timings are fitted with the observed
ones. The theoretical transit timings are derived from 
numerical orbital integrations of the involved star and planets. These calculations could be time-consuming and
the corresponding results could be very sensitive to 
some parameters in a highly non-linear way.
Therefore, the Levenberg-Marquardt minimization
algorithm \citep{Press1992} is often used
to search for better initial values in parameter spaces before MCMC sampling \citep{Hadden2016, Hadden2017}. This stage is very important and can make the
MCMC sampling process converge smoothly. 
However, details of the searching regions of parameter spaces are usually not described 
in astronomy literature. In addition, because the Levenberg-Marquardt minimization algorithm needs to perform numerical differentiation calculations, it is less stable than the Nelder-Mead algorithm \citep{NelderMead1965}.
In this paper, we propose to use   
Nelder-Mead algorithm as the first stage and 
MCMC sampling as the second stage in a
two-stage fitting process.

The $TTV$ plots are obtained and presented in 
Fig.\ref{fig:OC_org}, 
in which top panel is for the planet K2-266~d and 
bottom panel is for the planet K2-266~e.  The purple line in Fig. \ref{fig:OC_org} is from the best-fit model which will be described later.

Then, we employ the package {\it TTVFast}  \citep{Deck2014} to produce theoretical TTV signals
and fit to the observational counterparts. 
However, the modeled quantities have a highly non-linear dependence on governing parameters
and only can be obtained through a complicated numerical calculation, the fitting 
becomes a difficult optimization problem here.

Nelder-Mead algorithm \citep{NelderMead1965} is designed to solve the optimization problem of minimizing a given nonlinear function.
The method only requires function evaluations and does not require any derivative information, which makes it suitable for problems with non-smooth functions. 
The Nelder-Mead method is simplex-based and is introduced 
in \cite{Press1992} as a downhill simplex method in a multi-dimension space. 
A new implementation of Nelder-Mead algorithm followed \cite{GaoHan2010} in the package
{\it SciPy} \citep{Virtanen2020}
is used here. 
To start the processes in Nelder-Mead algorithm, the initial values of parameters in TTVFast, i.e. the mass $m$, 
orbital period $p$, orbital eccentricity $e$, argument of pericenter $w$, mean anomaly $Ma$, 
and orbital inclination $i$ for both planets are needed. 
Using the results in \cite{Rodri2018AJ} as a standard reference, 
a value of $m, p, e, w$ is randomly picked within the corresponding uncertainty ranges 
in Table 4 of \cite{Rodri2018AJ}, and a value of mean anomaly $Ma$ is randomly picked from $[0, 360]$. 
The inclinations of planet K2-266 d and e are set to be 89.46 degree and 89.45 degree
respectively, followed the values in \cite{Rodri2018AJ}. 
Repeating the above Nelder-Mead fitting process, which is summarized in Table \ref{table:NMfitting}, for 100 times, a best-fit solution with 
a chi-square $\chi^2$ value of 40.76 
is found with 17 degrees of freedom ($\chi^2_r=2.4$).

After we get this solution from the Nelder-Mead algorithm,
in order to obtain the uncertainties and
refine the solution, the MCMC sampling is employed \citep{Foreman2013}.
Similar to the MCMC process in  \cite{Rodri2018AJ},\cite{Lithwick2012}, 
\cite{Hadden2016}, and \cite{Hadden2017},
we consider to fit the mass, 
the orbital period, 
the orbital eccentricity, 
the argument of pericenter,
and the mean anomaly for both planets,
but fix their orbital inclinations.
Using $d$ and $e$ as the sub-indexes of variables associated to the planet K2-266~d and e, respectively. 
The fitting parameters are $m_d, 
\,m_e, \,p_d, \,p_e, \,e_d\cos w_d,  
\,e_d\sin w_d,
\,e_e\cos w_e, \,e_e\sin w_e,$ 
$Ma_d$ and $Ma_e$.
To simplify symbols, we define variables 
$ec_d\equiv e_d\cos w_d$,
$es_d\equiv e_d\sin w_d$, 
$ec_e\equiv e_e\cos w_e$, and 
$es_e\equiv e_e\sin w_e$. 
The prior distributions are uniform and no boundaries are set for all
fitting parameters during the MCMC sampling process, which is summarized in
Table \ref{table:MCMCfitting}. 

The posterior distributions of the above parameters of MCMC
sampling are presented in 
Fig.\ref{fig:fit10_corner}.
Then the MAP (maximum a posteriori) solution is set to be the result of the MCMC fitting process
\citep{Nascimbeni2024}, and the
standard deviations of the posterior distributions of the parameters are used as uncertainties. 
Table~\ref{table:bestMCMC10} gives the values of the best-fit parameter and the corresponding uncertainties of the above 10-parameter MCMC sampling process.

\begin{figure*}[!htb]
\hspace{-0.3cm}
\includegraphics[width=1.1\linewidth]{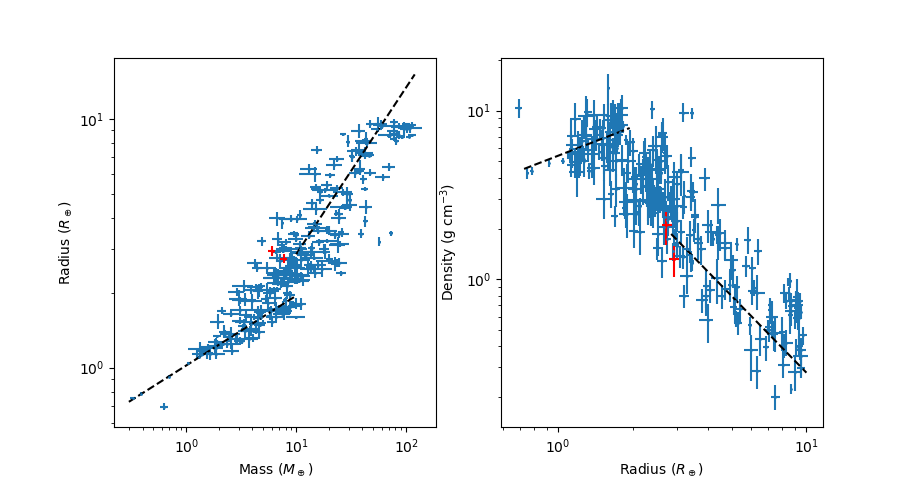}
\caption{Left Panel: exoplanets on the mass-radius
plane, where blue dots are the planets 
from PlanetS catalog
with mass less than 120 $M_\oplus$,
radius less than 10 $R_\oplus$, and
the left red dot is for the planet K2-266~d, the right red dot is for the planet K2-266~e,
dashed lines are adopted from \cite{Parc2024};
Right Panel: exoplanets on the radius-density
plane, where blue dots are exactly the same planets as in left panel,
the lower red dot is for 
the planet K2-266~d, the upper red dot is 
for the planet K2-266~e, dashed lines are based on the results in \cite{Parc2024}.
}
\label{figMR}
\end{figure*} 

\begin{figure*}[!htb]
\hspace{-0.5cm}
\includegraphics[width=1.05\linewidth]{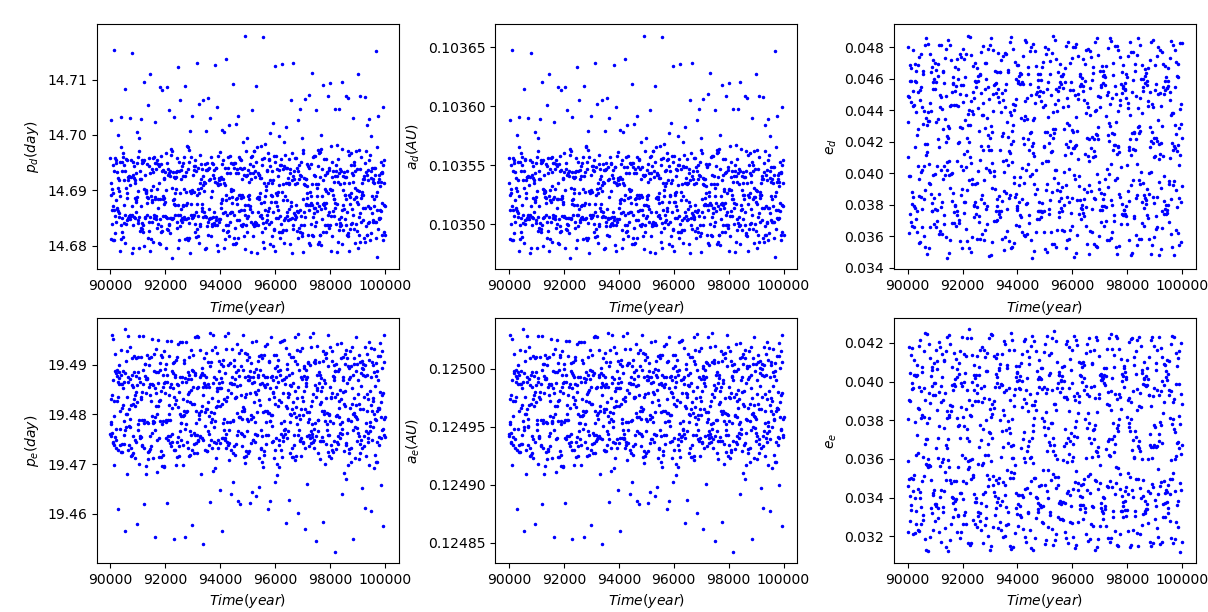}
\caption{The evolution of periods, semi-major axes, and orbital eccentricities of the best-fit solution during the final $10^{4}$ years of the simulation. The top panels are for the planet K2-266~d and the bottom panels are for the planet K2-266~e.}
\label{fig_n0tpae}
\end{figure*} 
\begin{figure*}[!htb]
\hspace{-0.5cm}
\includegraphics[width=1.1\linewidth]{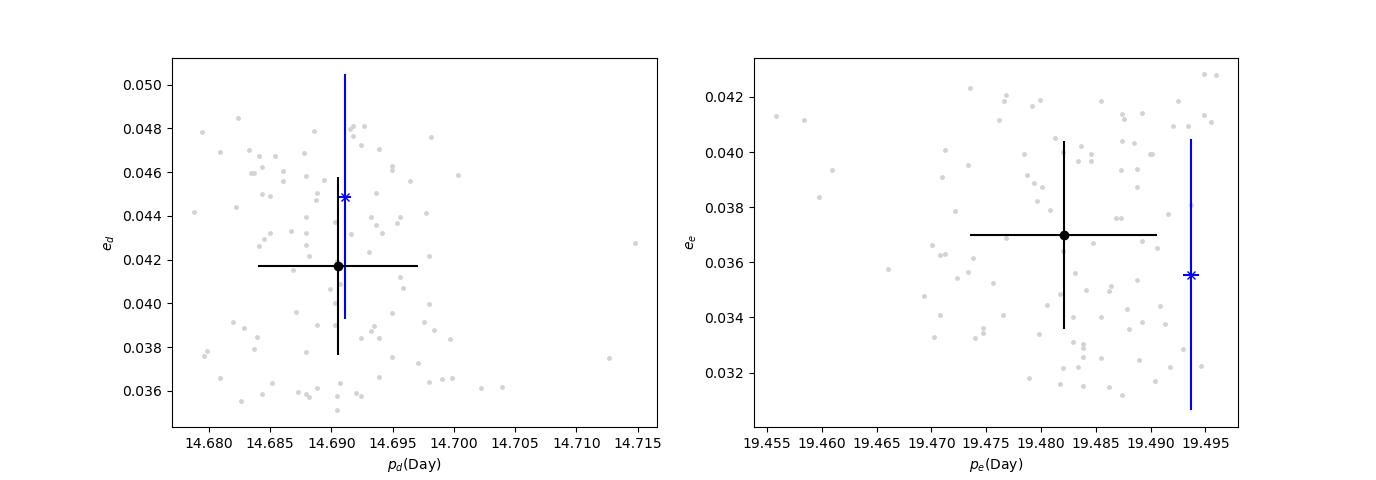}
\caption{The periods and eccentricities of the planet K2-266~d (left panel) and the planet K2-266~e (right panel) at 
the end of simulation $t=10^5$ years. The grey points are for the results of 100 simulations. The black points with error bars are for the means and standard deviations of these grey points, and
the blue points with error bars
are for the best-fit solutions and the corresponding uncertainties.} 
\label{figpe_final}
\end{figure*} 
\begin{figure}[!htb]
\hspace{-0.3cm}
\includegraphics[width=1.1\linewidth]{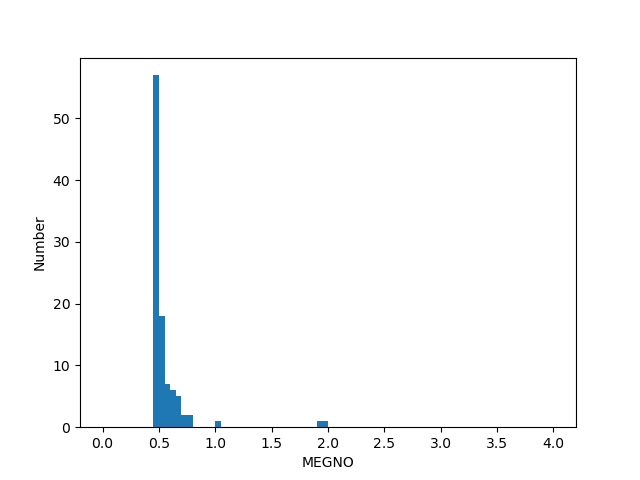}
\caption{The histogram of MEGNO indicators of 100
simulations.}
\label{figMEGNO}
\end{figure} 

The value of chi-square $\chi^2$ for the above best-fit solution is 40.02. With 27 fitting data points and 
ten fitting parameters, the degree of freedom is 
17. Thus, the value of reduced chi-square is 
$\chi^2_r= 2.35$.  
From the TTV plots in Fig.\ref{fig:OC_org}, it can be seen that the
first orange point in the top panel might be the one that contributes more to increase the
value of reduced chi-square. This point is derived from the CHEOPS transit light curve
of K2-266 d observed during Visit d1
(Table \ref{LogCHEOPS}).
A possible error could be caused by the lack
of data during the ingress and egress of that transit event.

\section{The Mass-Radius Relations}
\label{sec:mr}
Our long baseline TTV modeling  driven  by new CHEOPS and TESS data  leads to a set of updated 
parameters with higher precision 
(see Table \ref{table:bestMCMC10}).
Particularly, the precision of mass measurement is greatly improved.   
The impact and implication of this high-precision updated mass are presented here.
The mass could give hints about
the planetary formation scenario and the evolution process. It is also related to the orbital architecture of whole planetary system.

On the other hand, because planetary radii can be obtained from transit depths directly, the radius valley was discovered as a gap between sub-Neptunes and super-Earths \citep{Fulton2017AJ}. 
The relationship between the radius and mass of an exoplanet could provide constraints on its envelope fraction and internal structure. 
Therefore, it is important to study the mass-radius relations \citep{Otegi2020MR, 
Aguichine2021, Kubyshkina2022, Edmondson2023,
Muller2024, Ricard2024, Parc2024}, and through this relation,
different populations might be identified.

In order to address the radius valley and the 
mass-radius relation, 
\cite{Otegi2020MR} considered exoplanets with masses up to 120 $M_\oplus$ and only included
those with smaller mass and radius uncertainties.
Two populations, which correspond to 
the rocky and volatile-rich exoplanets, were found.  \cite{Edmondson2023} included Jupiters
into samples and identified three populations. Finally,
\cite{Parc2024} improved from the above two and provided 
an updated catalog, i.e. the PlanetS catalog, which is publicly available from the DACE platform\footnote{\url{https://dace.unige.ch/exoplanets}}.

Taken the values of planet radius from \cite{Rodri2018AJ}, the average density 
of planet K2-266~d is 
$1.32^{+0.26}_{-0.28}\ {\rm g cm^{-3}}$,
and the average density of planet K2-266~e is 
$2.09^{+0.41}_{-0.48}\  {\rm g cm^{-3}}$.
With these values of mass, radius, and average density, we can show their positions among the exoplanets of PlanetS catalog in the mass-radius diagram and also radius-density diagram, as presented in Fig. \ref{figMR}. 
Our K2-266~d and e are close to the boundary between rocky and volatile-rich populations, but
clearly belong to the volatile-rich population. 
In particular, the density of planet K2-266~d is 
lower than many volatile-rich planets. This interesting result might give hints for their
formation histories.

\section{Dynamical Simulations}
\label{sec:dynamics}
Through the two-stage fitting process with Nelder-Mead algorithm and MCMC sampling, the masses and orbital parameters of 
both planets K2-266~d and e
are finally determined. 
In this section, we perform dynamical simulations in order to study the 
orbital stability \citep{Petrovich2015, Kane2015,
Veras2017,  Liu2020,  Kovacs2022, Teixeira2023, Livesey2024}, 
regularity \citep{Volk2020, Hussain2020, Shevchenko2022}, and also the mean 
motion resonance between planets K2-266~d and e.

In order to have enough representative cases,
100 different initial conditions will be considered
and their corresponding results will be obtained
accordingly.
Following \citet{Becker2020AJ}, we consider
four planets, i.e. ignore those two planet candidates \citep{Rodri2018AJ}, in our simulations. 
The planetary masses and orbital elements of the planets K2-266~b and c are set to 
the values in \citet{Rodri2018AJ}.
As for the planet K2-266~d and e, the  
100 sets of planetary
masses and orbital parameters with smaller
chi-square values among the
MCMC samples are used.
During the simulations, the masses are fixed, and the orbital parameters are set as initial values and would vary as functions of time. Employing the orbit-integration code 
{\it REBOUND} \citep{ReinLiu2012}, 
these 100 dynamical simulations start from 
a time $t=t_{\rm ini}\equiv 0$ and extend to 
a time $t=t_{\rm end}\equiv 10^5$ years. 

In Fig.~\ref{fig_n0tpae},
we show the final $10^{4}$-year evolution of the simulation using the best-fit solution from Table\ref{table:bestMCMC10}
as the initial condition.
The upper panels are for planet K2-266~d, and the bottom panels are for planet K2-266~e. The orbital periods, 
semi-major axes, and orbital 
eccentricities are plotted as functions of time. 
They all oscillate around the initial values, so both planets'
future orbits derived from the best-fit solution 
are confirmed to be stable.

Then, the final snapshots of 100 simulations are presented
at the period-eccentricity plane in Fig.~\ref{figpe_final}. The left panel is for planet K2-266~d,
and the right panel is for planet K2-266~e. 
The grey points are the values of the periods and eccentricities at the time $t_{\rm end}$.
The black point with error bar are for the mean and standard deviation of these grey points, and
the blue point with error bar
are for the best-fit solution and the corresponding uncertainty. 
The majority of grey points shift to positions with slightly smaller periods but still quite close to the best-fit solution.

In addition, we also calculate the Mean Exponential Growth factor of Nearby Orbits (MEGNO) indicators \citep{Cincotta2003} for these 100 simulations. The chaotic orbits would have MEGNO indicators larger than 4. As presented in Fig.~\ref{figMEGNO}, all MEGNO indicators are much lower than 4, and most have values around 0.5. Thus, all these 100 orbits are regular, not chaotic.
The results show that these orbits are stable during our simulations. 

Moreover, from the period ratio, it is known that 
the planet K2-266~d and the planet K2-266~e are close
to 4:3 mean motion resonance.  \citet{Rodri2018AJ} showed that only a small fraction of their simulated
orbits are in true resonance. 
In order to examine this, 
we determine the values of 4:3 resonant argument
during the final $10^{4}$ years 
of the simulations. Among these 100 simulations,
none of their 4:3 resonant arguments are confined.
That is, there is no any libration case.
Thus, our results show that the 
planet K2-266~d and
the planet K2-266~e are close to, but not in 4:3 mean motion resonance.

\section{Conclusions}
\label{sec:conclusion}

With new CHEOPS and TESS data, the masses, orbital 
periods, and orbital eccentricities of the planet K2-266~d and the planet K2-266~e are updated, as presented in
Table \ref{table:bestMCMC10}. In particular, the precision of mass measurements has been raised to be one order of magnitude higher. With these new masses,
both planets K2-266~d and e are confirmed to belong to the volatile-rich population. Among these two, the planet K2-266~d 
has smaller density and is located in the outer parts of
the volatile-rich population in the mass-radius plane
and the radius-density plane shown in
Fig. \ref{figMR}.

With the above results, our dynamical simulations further confirm that 
the planetary orbits are stable and not chaotic.
In addition, the planet K2-266~d and
the planet K2-266~e are close to, but not in 4:3 mean motion resonance.
Finally, as TESS will re-observe the K2-266 system in Sector 89, it is expected that future observational data from various
facilities will contribute to the further  characterization and understanding of this system.
\clearpage

\section*{Acknowledgments}
We are grateful to the anonymous referee for many good suggestions, which greatly improve the presentation of this paper.
This work is supported by the grant from the National Science and Technology Council (NSTC), Taiwan. The grant numbers are NSTC 111-2112-M-007-035 and NSTC 113-2112-M-007-030. CHEOPS is an ESA mission in partnership with Switzerland with important contributions to the payload and the ground segment from Austria, Belgium, France, Germany, Hungary, Italy, Portugal, Spain, Sweden, and the United Kingdom. The CHEOPS Consortium gratefully acknowledges the support received by all the agencies, offices, universities, and industries involved. Their flexibility and willingness to explore new approaches were essential to the success of this mission. CHEOPS data analyzed in this paper are available in the CHEOPS mission archive via
\dataset[https://cheops.unige.ch/archive\_browser/]{https://cheops.unige.ch/archive_browser/}.
TESS data presented in this paper were obtained from the Mikulski Archive for Space Telescopes (MAST) at the Space Telescope Science Institute
(STScI). The specific
observations analyzed can be accessed via 
\dataset[https://doi.org/10.17909/t9-nmc8-f686]{https://doi.org/10.17909/t9-nmc8-f686} \citep{MAST2021}. 

\vspace{3mm}
\facilities{CHaracterising ExOPlanets Satellite (CHEOPS),
Transiting Exoplanet Survey Satellite (TESS)}

\software{
{\tt\string PyLightcurve} \citep{Tsiaras2016ApJ}, 
{\tt\string emcee} \citep{Foreman2013}, 
{\tt\string TTVFast} \citep{Deck2014},
{\tt\string SciPy} \citep{Virtanen2020}. }

\bibliography{K2266}{}

\begin{thebibliography}{}
\expandafter\ifx\csname natexlab\endcsname\relax\def\natexlab#1{#1}\fi
\providecommand{\url}[1]{\href{#1}{#1}}
\providecommand{\dodoi}[1]{doi:~\href{http://doi.org/#1}{\nolinkurl{#1}}}
\providecommand{\doeprint}[1]{\href{http://ascl.net/#1}{\nolinkurl{http://ascl.net/#1}}}
\providecommand{\doarXiv}[1]{\href{https://arxiv.org/abs/#1}{\nolinkurl{https://arxiv.org/abs/#1}}}

\bibitem[{{Aguichine} {et~al.}(2021){Aguichine}, {Mousis}, {Deleuil}, \& {Marcq}}]{Aguichine2021}
{Aguichine}, A., {Mousis}, O., {Deleuil}, M., \& {Marcq}, E. 2021, \apj, 914, 84, \dodoi{10.3847/1538-4357/abfa99}

\bibitem[{{Becker} {et~al.}(2020){Becker}, {Batygin}, {Fabrycky}, {Adams}, {Li}, {Vanderburg}, \& {Rodriguez}}]{Becker2020AJ}
{Becker}, J., {Batygin}, K., {Fabrycky}, D., {et~al.} 2020, \aj, 160, 254, \dodoi{10.3847/1538-3881/abbad3}

\bibitem[{{Benneke} {et~al.}(2019){Benneke}, {Wong}, {Piaulet}, {Knutson}, {Lothringer}, {Morley}, {Crossfield}, {Gao}, {Greene}, {Dressing}, {Dragomir}, {Howard}, {McCullough}, {Kempton}, {Fortney}, \& {Fraine}}]{Benneke2019}
{Benneke}, B., {Wong}, I., {Piaulet}, C., {et~al.} 2019, \apjl, 887, L14, \dodoi{10.3847/2041-8213/ab59dc}

\bibitem[{{Benz} {et~al.}(2021){Benz}, {Broeg}, {Fortier}, {Rando}, {Beck}, {Beck}, {Queloz}, {Ehrenreich}, {Maxted}, {Isaak}, {Billot}, {Alibert}, {Alonso}, {Ant{\'o}nio}, {Asquier}, {Bandy}, {B{\'a}rczy}, {Barrado}, {Barros}, {Baumjohann}, {Bekkelien}, {Bergomi}, {Biondi}, {Bonfils}, {Borsato}, {Brandeker}, {Busch}, {Cabrera}, {Cessa}, {Charnoz}, {Chazelas}, {Collier Cameron}, {Corral Van Damme}, {Cortes}, {Davies}, {Deleuil}, {Deline}, {Delrez}, {Demangeon}, {Demory}, {Erikson}, {Farinato}, {Fossati}, {Fridlund}, {Futyan}, {Gandolfi}, {Garcia Munoz}, {Gillon}, {Guterman}, {Gutierrez}, {Hasiba}, {Heng}, {Hernandez}, {Hoyer}, {Kiss}, {Kovacs}, {Kuntzer}, {Laskar}, {Lecavelier des Etangs}, {Lendl}, {L{\'o}pez}, {Lora}, {Lovis}, {L{\"u}ftinger}, {Magrin}, {Malvasio}, {Marafatto}, {Michaelis}, {de Miguel}, {Modrego}, {Munari}, {Nascimbeni}, {Olofsson}, {Ottacher}, {Ottensamer}, {Pagano}, {Palacios}, {Pall{\'e}}, {Peter}, {Piazza}, {Piotto}, {Pizarro}, {Pollaco}, {Ragazzoni}, {Ratti}, {Rauer}, {Ribas}, {Rieder},
  {Rohlfs}, {Safa}, {Salatti}, {Santos}, {Scandariato}, {S{\'e}gransan}, {Simon}, {Smith}, {Sordet}, {Sousa}, {Steller}, {Szab{\'o}}, {Szoke}, {Thomas}, {Tschentscher}, {Udry}, {Van Grootel}, {Viotto}, {Walter}, {Walton}, {Wildi}, \& {Wolter}}]{Benz2021}
{Benz}, W., {Broeg}, C., {Fortier}, A., {et~al.} 2021, Experimental Astronomy, 51, 109, \dodoi{10.1007/s10686-020-09679-4}

\bibitem[{{Borsato} {et~al.}(2024){Borsato}, {Degen}, {Leleu}, {Hooton}, {Egger}, {Bekkelien}, {Brandeker}, {Collier Cameron}, {G{\"u}nther}, {Nascimbeni}, {Persson}, {Bonfanti}, {Wilson}, {Correia}, {Zingales}, {Guillot}, {Triaud}, {Piotto}, {Gandolfi}, {Abe}, {Alibert}, {Alonso}, {B{\'a}rczy}, {Navascues}, {Barros}, {Baumjohann}, {Beck}, {Bendjoya}, {Benz}, {Billot}, {Broeg}, {Busch}, {Csizmadia}, {Cubillos}, {Davies}, {Deleuil}, {Deline}, {Delrez}, {Demangeon}, {Demory}, {Derekas}, {Edwards}, {Ehrenreich}, {Erikson}, {Fortier}, {Fossati}, {Fridlund}, {Gazeas}, {Gillon}, {G{\"u}del}, {Heitzmann}, {Helling}, {Hoyer}, {Isaak}, {Kiss}, {Korth}, {Lam}, {Laskar}, {Lecavelier des Etangs}, {Lendl}, {Magrin}, {Marafatto}, {Maxted}, {Mecina}, {M{\'e}karnia}, {Mordasini}, {Mura}, {Olofsson}, {Ottensamer}, {Pagano}, {Pall{\'e}}, {Peter}, {Pollacco}, {Queloz}, {Ragazzoni}, {Rando}, {Ratti}, {Rauer}, {Ribas}, {Salmon}, {Santos}, {Scandariato}, {S{\'e}gransan}, {Simon}, {Smith}, {Sousa}, {Stalport}, {Suarez}, {Sulis},
  {Szab{\'o}}, {Udry}, {Van Grootel}, {Venturini}, {Villaver}, {Walton}, \& {Wolter}}]{Borsato2024}
{Borsato}, L., {Degen}, D., {Leleu}, A., {et~al.} 2024, \aap, 689, A52, \dodoi{10.1051/0004-6361/202450974}

\bibitem[{{Borucki} {et~al.}(2010){Borucki}, {Koch}, {Basri}, {Batalha}, {Brown}, {Caldwell}, \& {Caldwell}}]{Borucki+2010Sci}
{Borucki}, W.~J., {Koch}, D., {Basri}, G., {et~al.} 2010, Science, 327, 977, \dodoi{10.1126/science.1185402}

\bibitem[{{Boujibar} {et~al.}(2020){Boujibar}, {Driscoll}, \& {Fei}}]{Boujibar2020}
{Boujibar}, A., {Driscoll}, P., \& {Fei}, Y. 2020, Journal of Geophysical Research (Planets), 125, e06124, \dodoi{10.1029/2019JE006124}

\bibitem[{{Cincotta} {et~al.}(2003){Cincotta}, {Giordano}, \& {Sim{\'o}}}]{Cincotta2003}
{Cincotta}, P.~M., {Giordano}, C.~M., \& {Sim{\'o}}, C. 2003, Physica D Nonlinear Phenomena, 182, 151, \dodoi{10.1016/S0167-2789(03)00103-9}

\bibitem[{{Claret} \& {Southworth}(2023)}]{claret}
{Claret}, A., \& {Southworth}, J. 2023, \aap, 674, A63, \dodoi{10.1051/0004-6361/202346478}

\bibitem[{{Deck} {et~al.}(2014){Deck}, {Agol}, {Holman}, \& {Nesvorn{\'y}}}]{Deck2014}
{Deck}, K.~M., {Agol}, E., {Holman}, M.~J., \& {Nesvorn{\'y}}, D. 2014, \apj, 787, 132, \dodoi{10.1088/0004-637X/787/2/132}

\bibitem[{{Deline} {et~al.}(2022){Deline}, {Hooton}, {Lendl}, {Morris}, {Salmon}, {Olofsson}, {Broeg}, {Ehrenreich}, {Beck}, {Brandeker}, {Hoyer}, {Sulis}, {Van Grootel}, {Bourrier}, {Demangeon}, {Demory}, {Heng}, {Parviainen}, {Serrano}, {Singh}, {Bonfanti}, {Fossati}, {Kitzmann}, {Sousa}, {Wilson}, {Alibert}, {Alonso}, {Anglada}, {B{\'a}rczy}, {Barrado Navascues}, {Barros}, {Baumjohann}, {Beck}, {Bekkelien}, {Benz}, {Billot}, {Bonfils}, {Cabrera}, {Charnoz}, {Collier Cameron}, {Corral van Damme}, {Csizmadia}, {Davies}, {Deleuil}, {Delrez}, {de Roche}, {Erikson}, {Fortier}, {Fridlund}, {Futyan}, {Gandolfi}, {Gillon}, {G{\"u}del}, {Gutermann}, {Hasiba}, {Isaak}, {Kiss}, {Laskar}, {Lecavelier des Etangs}, {Lovis}, {Magrin}, {Maxted}, {Munari}, {Nascimbeni}, {Ottensamer}, {Pagano}, {Pall{\'e}}, {Peter}, {Piotto}, {Pollacco}, {Queloz}, {Ragazzoni}, {Rando}, {Rauer}, {Ribas}, {Santos}, {Scandariato}, {S{\'e}gransan}, {Simon}, {Smith}, {Steller}, {Szab{\'o}}, {Thomas}, {Udry}, {Walter}, \& {Walton}}]{Deline2022}
{Deline}, A., {Hooton}, M.~J., {Lendl}, M., {et~al.} 2022, \aap, 659, A74, \dodoi{10.1051/0004-6361/202142400}

\bibitem[{{Dorn} {et~al.}(2015){Dorn}, {Khan}, {Heng}, {Connolly}, {Alibert}, {Benz}, \& {Tackley}}]{Dorn2015}
{Dorn}, C., {Khan}, A., {Heng}, K., {et~al.} 2015, \aap, 577, A83, \dodoi{10.1051/0004-6361/201424915}

\bibitem[{{Eastman} {et~al.}(2010){Eastman}, {Siverd}, \& {Gaudi}}]{Eastman2010}
{Eastman}, J., {Siverd}, R., \& {Gaudi}, B.~S. 2010, \pasp, 122, 935, \dodoi{10.1086/655938}

\bibitem[{{Edmondson} {et~al.}(2023){Edmondson}, {Norris}, \& {Kerins}}]{Edmondson2023}
{Edmondson}, K., {Norris}, J., \& {Kerins}, E. 2023, arXiv e-prints, arXiv:2310.16733, \dodoi{10.48550/arXiv.2310.16733}

\bibitem[{{Egger} {et~al.}(2024){Egger}, {Osborn}, {Kubyshkina}, {Mordasini}, {Alibert}, {G{\"u}nther}, {Lendl}, {Brandeker}, {Heitzmann}, {Leleu}, {Damasso}, {Bonfanti}, {Wilson}, {Sousa}, {Haldemann}, {Delrez}, {Hooton}, {Zingales}, {Luque}, {Alonso}, {Asquier}, {B{\'a}rczy}, {Navascues}, {Barros}, {Baumjohann}, {Benz}, {Billot}, {Borsato}, {Broeg}, {Buder}, {Castro-Gonz{\'a}lez}, {Cameron}, {Correia}, {Cortes}, {Csizmadia}, {Cubillos}, {Davies}, {Deleuil}, {Deline}, {Demangeon}, {Demory}, {Derekas}, {Edwards}, {Ehrenreich}, {Erikson}, {Fortier}, {Fossati}, {Fridlund}, {Gandolfi}, {Gazeas}, {Gillon}, {G{\"u}del}, {Helling}, {Isaak}, {Kiss}, {Korth}, {Lam}, {Laskar}, {Lavie}, {des Etangs}, {Lovis}, {Luntzer}, {Magrin}, {Maxted}, {Mer{\'\i}n}, {Munari}, {Nascimbeni}, {Olofsson}, {Ottensamer}, {Pagano}, {Pall{\'e}}, {Peter}, {Piazza}, {Piotto}, {Pollacco}, {Queloz}, {Ragazzoni}, {Rando}, {Rauer}, {Ribas}, {Rodrigues}, {Santos}, {Scandariato}, {S{\'e}gransan}, {Simon}, {Smith}, {Stalport}, {Sulis}, {Szab{\'o}},
  {Udry}, {Van Grootel}, {Venturini}, {Villaver}, \& {Walton}}]{Egger2024}
{Egger}, J.~A., {Osborn}, H.~P., {Kubyshkina}, D., {et~al.} 2024, \aap, 688, A223, \dodoi{10.1051/0004-6361/202450472}

\bibitem[{{Foreman-Mackey} {et~al.}(2013){Foreman-Mackey}, {Hogg}, {Lang}, \& {Goodman}}]{Foreman2013}
{Foreman-Mackey}, D., {Hogg}, D.~W., {Lang}, D., \& {Goodman}, J. 2013, \pasp, 125, 306, \dodoi{10.1086/670067}

\bibitem[{{Fridlund} {et~al.}(2024){Fridlund}, {Georgieva}, {Bonfanti}, {Alibert}, {Persson}, {Gandolfi}, {Beck}, {Deline}, {Hoyer}, {Olofsson}, {Wilson}, {Barrag{\'a}n}, {Fossati}, {Mustill}, {Brandeker}, {Hatzes}, {Flor{\'e}n}, {Simola}, {Hooton}, {Luque}, {Sousa}, {Egger}, {Antoniadis-Karnavas}, {Salmon}, {Adibekyan}, {Alonso}, {Anglada}, {B{\'a}rczy}, {Barrado Navascues}, {Barros}, {Baumjohann}, {Beck}, {Benz}, {Bonfils}, {Broeg}, {Cabrera}, {Charnoz}, {Collier Cameron}, {Csizmadia}, {Davies}, {Deeg}, {Deleuil}, {Delrez}, {Demangeon}, {Demory}, {Ehrenreich}, {Erikson}, {Esposito}, {Fortier}, {Gillon}, {G{\"u}del}, {Heng}, {Isaak}, {Kiss}, {Korth}, {Laskar}, {Lecavelier des Etangs}, {Lendl}, {Livingston}, {Lovis}, {Magrin}, {Maxted}, {Muresan}, {Nascimbeni}, {Ottensamer}, {Pagano}, {Pall{\'e}}, {Peter}, {Piotto}, {Pollacco}, {Queloz}, {Ragazzoni}, {Rando}, {Rauer}, {Redfield}, {Ribas}, {Santos}, {Scandariato}, {S{\'e}gransan}, {Serrano}, {Simon}, {Smith}, {Steller}, {Szab{\'o}}, {Thomas}, {Udry}, {Van
  Eylen}, {Van Grootel}, \& {Walton}}]{Fridlund2024}
{Fridlund}, M., {Georgieva}, I.~Y., {Bonfanti}, A., {et~al.} 2024, \aap, 684, A12, \dodoi{10.1051/0004-6361/202243838}

\bibitem[{{Fulton} {et~al.}(2017){Fulton}, {Petigura}, {Howard}, {Isaacson}, {Marcy}, {Cargile}, {Hebb}, {Weiss}, {Johnson}, {Morton}, {Sinukoff}, {Crossfield}, \& {Hirsch}}]{Fulton2017AJ}
{Fulton}, B.~J., {Petigura}, E.~A., {Howard}, A.~W., {et~al.} 2017, \aj, 154, 109, \dodoi{10.3847/1538-3881/aa80eb}

\bibitem[{{Gajendran} {et~al.}(2024){Gajendran}, {Jiang}, {Yeh}, \& {Sariya}}]{Gajendran2024}
{Gajendran}, S., {Jiang}, I.-G., {Yeh}, L.-C., \& {Sariya}, D.~P. 2024, \mnras, 528, 7202, \dodoi{10.1093/mnras/stae501}

\bibitem[{{Gao} \& {Han}(2010)}]{GaoHan2010}
{Gao}, F., \& {Han}, L. 2010, Computational Optimization and Applications, 51, 259

\bibitem[{{Hadden} \& {Lithwick}(2016)}]{Hadden2016}
{Hadden}, S., \& {Lithwick}, Y. 2016, \apj, 828, 44, \dodoi{10.3847/0004-637X/828/1/44}

\bibitem[{{Hadden} \& {Lithwick}(2017)}]{Hadden2017}
---. 2017, \aj, 154, 5, \dodoi{10.3847/1538-3881/aa71ef}

\bibitem[{{Hirano} {et~al.}(2021){Hirano}, {Livingston}, {Fukui}, {Narita}, {Harakawa}, {Ishikawa}, {Miyakawa}, {Kimura}, {Nakayama}, {Fujita}, {Hori}, {Stassun}, {Bieryla}, {Cadieux}, {Ciardi}, {Collins}, {Ikoma}, {Vanderburg}, {Barclay}, {Brasseur}, {de Leon}, {Doty}, {Doyon}, {Esparza-Borges}, {Esquerdo}, {Furlan}, {Gaidos}, {Gonzales}, {Hodapp}, {Howell}, {Isogai}, {Jacobson}, {Jenkins}, {Jensen}, {Kawauchi}, {Kotani}, {Kudo}, {Kurita}, {Kurokawa}, {Kusakabe}, {Kuzuhara}, {Lafreni{\`e}re}, {Latham}, {Massey}, {Mori}, {Murgas}, {Nishikawa}, {Nishiumi}, {Omiya}, {Paegert}, {Palle}, {Parviainen}, {Quinn}, {Ricker}, {Schwarz}, {Seager}, {Tamura}, {Tenenbaum}, {Terada}, {Vanderspek}, {Vievard}, {Watanabe}, \& {Winn}}]{Hirano2021}
{Hirano}, T., {Livingston}, J.~H., {Fukui}, A., {et~al.} 2021, \aj, 162, 161, \dodoi{10.3847/1538-3881/ac0fdc}

\bibitem[{{Howell} {et~al.}(2014){Howell}, {Sobeck}, {Haas}, {Still}, {Barclay}, {Mullally}, {Troeltzsch}, {Aigrain}, {Bryson}, {Caldwell}, {Chaplin}, {Cochran}, {Huber}, {Marcy}, {Miglio}, {Najita}, {Smith}, {Twicken}, \& {Fortney}}]{Howell2014PASP}
{Howell}, S.~B., {Sobeck}, C., {Haas}, M., {et~al.} 2014, \pasp, 126, 398, \dodoi{10.1086/676406}

\bibitem[{{Hoyer} {et~al.}(2020){Hoyer}, {Guterman}, {Demangeon}, {Sousa}, {Deleuil}, {Meunier}, \& {Benz}}]{Hoyer2020}
{Hoyer}, S., {Guterman}, P., {Demangeon}, O., {et~al.} 2020, \aap, 635, A24, \dodoi{10.1051/0004-6361/201936325}

\bibitem[{{Hussain} \& {Tamayo}(2020)}]{Hussain2020}
{Hussain}, N., \& {Tamayo}, D. 2020, \mnras, 491, 5258, \dodoi{10.1093/mnras/stz3402}

\bibitem[{{Jenkins} {et~al.}(2016){Jenkins}, {Twicken}, {McCauliff}, {Campbell}, {Sanderfer}, {Lung}, {Mansouri-Samani}, {Girouard}, {Tenenbaum}, {Klaus}, {Smith}, {Caldwell}, {Chacon}, {Henze}, {Heiges}, {Latham}, {Morgan}, {Swade}, {Rinehart}, \& {Vanderspek}}]{Jenkins2016}
{Jenkins}, J.~M., {Twicken}, J.~D., {McCauliff}, S., {et~al.} 2016, in Society of Photo-Optical Instrumentation Engineers (SPIE) Conference Series, Vol. 9913, Software and Cyberinfrastructure for Astronomy IV, ed. G.~{Chiozzi} \& J.~C. {Guzman}, 99133E, \dodoi{10.1117/12.2233418}

\bibitem[{{Kane}(2015)}]{Kane2015}
{Kane}, S.~R. 2015, \apjl, 814, L9, \dodoi{10.1088/2041-8205/814/1/L9}

\bibitem[{{Kite} {et~al.}(2019){Kite}, {Fegley}, {Schaefer}, \& {Ford}}]{Kite2019}
{Kite}, E.~S., {Fegley}, Bruce, J., {Schaefer}, L., \& {Ford}, E.~B. 2019, \apjl, 887, L33, \dodoi{10.3847/2041-8213/ab59d9}

\bibitem[{{Koch} {et~al.}(2010){Koch}, {Borucki}, {Basri}, {Batalha}, {Brown}, \& {Caldwell}}]{Koch2010}
{Koch}, D.~G., {Borucki}, W.~J., {Basri}, G., {et~al.} 2010, \apjl, 713, L79, \dodoi{10.1088/2041-8205/713/2/L79}

\bibitem[{{Kov{\'a}cs} {et~al.}(2022){Kov{\'a}cs}, {Pszota}, {K{\H{o}}v{\'a}ri}, {Forg{\'a}cs-Dajka}, \& {S{\'a}ndor}}]{Kovacs2022}
{Kov{\'a}cs}, T., {Pszota}, M., {K{\H{o}}v{\'a}ri}, E., {Forg{\'a}cs-Dajka}, E., \& {S{\'a}ndor}, Z. 2022, \mnras, 517, 5160, \dodoi{10.1093/mnras/stac3010}

\bibitem[{{Kubyshkina} \& {Fossati}(2022)}]{Kubyshkina2022}
{Kubyshkina}, D., \& {Fossati}, L. 2022, \aap, 668, A178, \dodoi{10.1051/0004-6361/202244916}

\bibitem[{{Kubyshkina} {et~al.}(2019){Kubyshkina}, {Cubillos}, {Fossati}, {Erkaev}, {Johnstone}, {Kislyakova}, {Lammer}, {Lendl}, {Odert}, \& {G{\"u}del}}]{Kubysh2019}
{Kubyshkina}, D., {Cubillos}, P.~E., {Fossati}, L., {et~al.} 2019, \apj, 879, 26, \dodoi{10.3847/1538-4357/ab1e42}

\bibitem[{{Lammer} {et~al.}(2016){Lammer}, {Erkaev}, {Fossati}, {Juvan}, {Odert}, {Cubillos}, {Guenther}, {Kislyakova}, {Johnstone}, {L{\"u}ftinger}, \& {G{\"u}del}}]{Lammer2016}
{Lammer}, H., {Erkaev}, N.~V., {Fossati}, L., {et~al.} 2016, \mnras, 461, L62, \dodoi{10.1093/mnrasl/slw095}

\bibitem[{{Latham} {et~al.}(2011){Latham}, {Rowe}, {Quinn}, {Batalha}, {Borucki}, {Brown}, {Bryson}, {Buchhave}, {Caldwell}, {Carter}, {Christiansen}, {Ciardi}, {Cochran}, {Dunham}, {Fabrycky}, {Ford}, {Gautier}, {Gilliland}, {Holman}, {Howell}, {Ibrahim}, {Isaacson}, {Jenkins}, {Koch}, {Lissauer}, {Marcy}, {Quintana}, {Ragozzine}, {Sasselov}, {Shporer}, {Steffen}, {Welsh}, \& {Wohler}}]{Latham2011}
{Latham}, D.~W., {Rowe}, J.~F., {Quinn}, S.~N., {et~al.} 2011, \apjl, 732, L24, \dodoi{10.1088/2041-8205/732/2/L24}

\bibitem[{{Levi} {et~al.}(2014){Levi}, {Sasselov}, \& {Podolak}}]{Levi2014}
{Levi}, A., {Sasselov}, D., \& {Podolak}, M. 2014, \apj, 792, 125, \dodoi{10.1088/0004-637X/792/2/125}

\bibitem[{{Lithwick} {et~al.}(2012){Lithwick}, {Xie}, \& {Wu}}]{Lithwick2012}
{Lithwick}, Y., {Xie}, J., \& {Wu}, Y. 2012, \apj, 761, 122, \dodoi{10.1088/0004-637X/761/2/122}

\bibitem[{{Liu} {et~al.}(2020){Liu}, {Gong}, \& {Li}}]{Liu2020}
{Liu}, C., {Gong}, S.-P., \& {Li}, J.-F. 2020, Research in Astronomy and Astrophysics, 20, 144, \dodoi{10.1088/1674-4527/20/9/144}

\bibitem[{{Livesey} {et~al.}(2024){Livesey}, {Barnes}, \& {Deitrick}}]{Livesey2024}
{Livesey}, J.~R., {Barnes}, R., \& {Deitrick}, R. 2024, \apj, 964, 4, \dodoi{10.3847/1538-4357/ad1ff4}

\bibitem[{{Lopez} \& {Fortney}(2014)}]{Lopez2014}
{Lopez}, E.~D., \& {Fortney}, J.~J. 2014, \apj, 792, 1, \dodoi{10.1088/0004-637X/792/1/1}

\bibitem[{{Luque} {et~al.}(2023){Luque}, {Osborn}, {Leleu}, {Pall{\'e}}, {Bonfanti}, \& {Barrag{\'a}n}}]{Luque2023}
{Luque}, R., {Osborn}, H.~P., {Leleu}, A., {et~al.} 2023, \nat, 623, 932, \dodoi{10.1038/s41586-023-06692-3}

\bibitem[{{MAST Data Products}(2021)}]{MAST2021}
{MAST Data Products}. 2021, TESS Light Curves - All Sectors,  STScI/MAST, \dodoi{10.17909/T9-NMC8-F686}

\bibitem[{{Maxted} {et~al.}(2022){Maxted}, {Ehrenreich}, {Wilson}, {Alibert}, {Cameron}, {Hoyer}, {Sousa}, {Olofsson}, {Bekkelien}, {Deline}, {Delrez}, {Bonfanti}, {Borsato}, {Alonso}, {Anglada Escud{\'e}}, {Barrado}, {Barros}, {Baumjohann}, {Beck}, {Beck}, {Benz}, {Billot}, {Biondi}, {Bonfils}, {Brandeker}, {Broeg}, {B{\'a}rczy}, {Cabrera}, {Charnoz}, {Corral Van Damme}, {Csizmadia}, {Davies}, {Deleuil}, {Demangeon}, {Demory}, {Erikson}, {Flor{\'e}n}, {Fortier}, {Fossati}, {Fridlund}, {Futyan}, {Gandolfi}, {Gillon}, {Guedel}, {Guterman}, {Heng}, {Isaak}, {Kiss}, {Laskar}, {Lecavelier des Etangs}, {Lendl}, {Lovis}, {Magrin}, {Nascimbeni}, {Ottensamer}, {Pagano}, {Pall{\'e}}, {Peter}, {Piotto}, {Pollacco}, {Pozuelos}, {Queloz}, {Ragazzoni}, {Rando}, {Rauer}, {Reimers}, {Ribas}, {Salmon}, {Santos}, {Scandariato}, {Simon}, {Smith}, {Steller}, {Swayne}, {Szab{\'o}}, {S{\'e}gransan}, {Thomas}, {Udry}, {Van Grootel}, \& {Walton}}]{maxted_pycheops}
{Maxted}, P.~F.~L., {Ehrenreich}, D., {Wilson}, T.~G., {et~al.} 2022, \mnras, 514, 77, \dodoi{10.1093/mnras/stab3371}

\bibitem[{{Moore} {et~al.}(2024){Moore}, {David}, {Zhang}, \& {Cowan}}]{Moore2024}
{Moore}, K., {David}, B., {Zhang}, A.~Y., \& {Cowan}, N.~B. 2024, \apj, 972, 131, \dodoi{10.3847/1538-4357/ad6444}

\bibitem[{{Morello} {et~al.}(2020){Morello}, {Claret}, {Martin-Lagarde}, {Cossou}, {Tsiaras}, \& {Lagage}}]{morello_exotethys}
{Morello}, G., {Claret}, A., {Martin-Lagarde}, M., {et~al.} 2020, \aj, 159, 75, \dodoi{10.3847/1538-3881/ab63dc}

\bibitem[{{Morris} {et~al.}(2021){Morris}, {Delrez}, {Brandeker}, {Cameron}, {Simon}, {Futyan}, {Olofsson}, {Hoyer}, {Fortier}, {Demory}, {Lendl}, {Wilson}, {Oshagh}, {Heng}, {Ehrenreich}, {Sulis}, {Alibert}, {Alonso}, {Anglada Escud{\'e}}, {Barrado}, {Barros}, {Baumjohann}, {Beck}, {Beck}, {Bekkelien}, {Benz}, {Bergomi}, {Billot}, {Bonfils}, {Bourrier}, {Broeg}, {B{\'a}rczy}, {Cabrera}, {Charnoz}, {Davies}, {De Miguel Ferreras}, {Deleuil}, {Deline}, {Demangeon}, {Erikson}, {Floren}, {Fossati}, {Fridlund}, {Gandolfi}, {Garc{\'\i}a Mu{\~n}oz}, {Gillon}, {Guedel}, {Guterman}, {Isaak}, {Kiss}, {Laskar}, {Lecavelier des Etangs}, {Lieder}, {Lovis}, {Magrin}, {Maxted}, {Nascimbeni}, {Ottensamer}, {Pagano}, {Pall{\'e}}, {Peter}, {Piotto}, {Pizarro Rubio}, {Pollacco}, {Pozuelos}, {Queloz}, {Ragazzoni}, {Rando}, {Rauer}, {Ribas}, {Santos}, {Scandariato}, {Smith}, {Sousa}, {Steller}, {Szab{\'o}}, {S{\'e}gransan}, {Thomas}, {Udry}, {Ulmer}, {Van Grootel}, \& {Walton}}]{Morris2021}
{Morris}, B.~M., {Delrez}, L., {Brandeker}, A., {et~al.} 2021, \aap, 653, A173, \dodoi{10.1051/0004-6361/202140892}

\bibitem[{{Mortier} {et~al.}(2016){Mortier}, {Faria}, {Santos}, {Rajpaul}, {Figueira}, {Boisse}, {Collier Cameron}, {Dumusque}, {Lo Curto}, {Lovis}, {Mayor}, {Melo}, {Pepe}, {Queloz}, \& {Santerne}}]{Mortier2016}
{Mortier}, A., {Faria}, J.~P., {Santos}, N.~C., {et~al.} 2016, \aap, 585, A135, \dodoi{10.1051/0004-6361/201526905}

\bibitem[{{M{\"u}ller} {et~al.}(2024){M{\"u}ller}, {Baron}, {Helled}, {Bouchy}, \& {Parc}}]{Muller2024}
{M{\"u}ller}, S., {Baron}, J., {Helled}, R., {Bouchy}, F., \& {Parc}, L. 2024, \aap, 686, A296, \dodoi{10.1051/0004-6361/202348690}

\bibitem[{{Nascimbeni} {et~al.}(2023){Nascimbeni}, {Borsato}, {Zingales}, {Piotto}, {Pagano}, {Beck}, {Broeg}, {Ehrenreich}, {Hoyer}, {Majidi}, {Granata}, {Sousa}, {Wilson}, {Van Grootel}, {Bonfanti}, {Salmon}, {Mustill}, {Delrez}, {Alibert}, {Alonso}, {Anglada}, {B{\'a}rczy}, {Barrado}, {Barros}, {Baumjohann}, {Beck}, {Benz}, {Bergomi}, {Billot}, {Bonfils}, {Brandeker}, {Cabrera}, {Charnoz}, {Collier Cameron}, {Csizmadia}, {Cubillos}, {Davies}, {Deleuil}, {Deline}, {Demangeon}, {Demory}, {Erikson}, {Fortier}, {Fossati}, {Fridlund}, {Gandolfi}, {Gillon}, {G{\"u}del}, {Isaak}, {Kiss}, {Laskar}, {Lecavelier des Etangs}, {Lendl}, {Lovis}, {Luque}, {Magrin}, {Maxted}, {Mordasini}, {Olofsson}, {Ottensamer}, {Pall{\'e}}, {Peter}, {Piazza}, {Pollacco}, {Queloz}, {Ragazzoni}, {Rando}, {Ratti}, {Rauer}, {Ribas}, {Santos}, {Scandariato}, {S{\'e}gransan}, {Simon}, {Smith}, {Steinberger}, {Steller}, {Szab{\'o}}, {Thomas}, {Udry}, {Venturini}, {Walton}, \& {Wolter}}]{Nasci2023}
{Nascimbeni}, V., {Borsato}, L., {Zingales}, T., {et~al.} 2023, \aap, 673, A42, \dodoi{10.1051/0004-6361/202245486}

\bibitem[{{Nascimbeni} {et~al.}(2024){Nascimbeni}, {Borsato}, {Leonardi}, {Sousa}, {Wilson}, {Fortier}, {Heitzmann}, {Mantovan}, {Luque}, {Zingales}, {Piotto}, {Alibert}, {Alonso}, {B{\'a}rczy}, {Barrado Navascues}, {Barros}, {Baumjohann}, {Beck}, {Benz}, {Billot}, {Biondi}, {Brandeker}, {Broeg}, {Busch}, {Collier Cameron}, {Correia}, {Csizmadia}, {Cubillos}, {Davies}, {Deleuil}, {Deline}, {Delrez}, {Demangeon}, {Demory}, {Derekas}, {Edwards}, {Ehrenreich}, {Erikson}, {Fossati}, {Fridlund}, {Gandolfi}, {Gazeas}, {Gillon}, {G{\"u}del}, {G{\"u}nther}, {Helling}, {Isaak}, {Kerschbaum}, {Kiss}, {Korth}, {Lam}, {Laskar}, {Lecavelier des Etangs}, {Leleu}, {Lendl}, {Magrin}, {Maxted}, {Mer{\'\i}n}, {Mordasini}, {Olofsson}, {Ottensamer}, {Pagano}, {Pall{\'e}}, {Peter}, {Pollacco}, {Queloz}, {Ragazzoni}, {Rando}, {Rauer}, {Ribas}, {Santos}, {Scandariato}, {S{\'e}gransan}, {Simon}, {Smith}, {Southworth}, {Stalport}, {Sulis}, {Szab{\'o}}, {Udry}, {Ulmer}, {Van Grootel}, {Venturini}, {Villaver}, \&
  {Walton}}]{Nascimbeni2024}
{Nascimbeni}, V., {Borsato}, L., {Leonardi}, P., {et~al.} 2024, \aap, 690, A349, \dodoi{10.1051/0004-6361/202450852}

\bibitem[{{Nelder} \& {Mead}(1965)}]{NelderMead1965}
{Nelder}, J.~A., \& {Mead}, R. 1965, Computer Journal, 7, 308

\bibitem[{{Nixon} \& {Madhusudhan}(2021)}]{Nixon2021}
{Nixon}, M.~C., \& {Madhusudhan}, N. 2021, \mnras, 505, 3414, \dodoi{10.1093/mnras/stab1500}

\bibitem[{{Osborn} {et~al.}(2023){Osborn}, {Nowak}, {H{\'e}brard}, {Masseron}, {Lillo-Box}, {Pall{\'e}}, {Bekkelien}, {Flor{\'e}n}, {Guterman}, {Simon}, {Adibekyan}, {Bieryla}, {Borsato}, {Brandeker}, {Ciardi}, {Collier Cameron}, {Collins}, {Egger}, {Gandolfi}, {Hooton}, {Latham}, {Lendl}, {Matthews}, {Tuson}, {Ulmer-Moll}, {Vanderburg}, {Wilson}, {Ziegler}, {Alibert}, {Alonso}, {Anglada}, {Arnold}, {Asquier}, {Barrado y Navascues}, {Baumjohann}, {Beck}, {Belinski}, {Benz}, {Biondi}, {Boisse}, {Bonfils}, {Broeg}, {Buchhave}, {B{\'a}rczy}, {Barros}, {Cabrera}, {Cardona Guillen}, {Carleo}, {Castro-Gonz{\'a}lez}, {Charnoz}, {Christiansen}, {Cortes-Zuleta}, {Csizmadia}, {Dalal}, {Davies}, {Deleuil}, {Delfosse}, {Delrez}, {Demory}, {Dunlavey}, {Ehrenreich}, {Erikson}, {Fernandes}, {Fortier}, {Forveille}, {Fossati}, {Fridlund}, {Gillon}, {Goeke}, {Goliguzova}, {Gonzales}, {G{\"u}nther}, {G{\"u}del}, {Heidari}, {Henze}, {Howell}, {Hoyer}, {Frey}, {Isaak}, {Jenkins}, {Kiefer}, {Kiss}, {Korth}, {Maxted}, {Laskar},
  {Lecavelier des Etangs}, {Lovis}, {Lund}, {Luque}, {Magrin}, {Almenara}, {Martioli}, {Mecina}, {Medina}, {Moldovan}, {Morales-Calder{\'o}n}, {Morello}, {Moutou}, {Murgas}, {Jensen}, {Nascimbeni}, {Olofsson}, {Ottensamer}, {Pagano}, {Peter}, {Piotto}, {Pollacco}, {Queloz}, {Ragazzoni}, {Rando}, {Rauer}, {Ribas}, {Ricker}, {Demangeon}, {Smith}, {Santos}, {Scandariato}, {Seager}, {Sousa}, {Steller}, {Szab{\'o}}, {S{\'e}gransan}, {Thomas}, {Udry}, {Ulmer}, {Van Grootel}, {Vanderspek}, {Walton}, \& {Winn}}]{osborn_hip9618}
{Osborn}, H.~P., {Nowak}, G., {H{\'e}brard}, G., {et~al.} 2023, \mnras, 523, 3069, \dodoi{10.1093/mnras/stad1319}

\bibitem[{{Otegi} {et~al.}(2020){Otegi}, {Bouchy}, \& {Helled}}]{Otegi2020MR}
{Otegi}, J.~F., {Bouchy}, F., \& {Helled}, R. 2020, \aap, 634, A43, \dodoi{10.1051/0004-6361/201936482}

\bibitem[{{Pagano} {et~al.}(2024){Pagano}, {Scandariato}, {Singh}, {Lendl}, {Queloz}, {Simon}, {Sousa}, {Brandeker}, {Cameron}, {Sulis}, {Van Grootel}, {Wilson}, {Alibert}, {Alonso}, {Anglada}, {B{\'a}rczy}, {Navascues}, {Barros}, {Baumjohann}, {Beck}, {Beck}, {Benz}, {Billot}, {Bonfils}, {Borsato}, {Broeg}, {Bruno}, {Carone}, {Charnoz}, {Corral van Damme}, {Csizmadia}, {Cubillos}, {Davies}, {Deleuil}, {Deline}, {Delrez}, {Demangeon}, {Demory}, {Ehrenreich}, {Erikson}, {Fortier}, {Fossati}, {Fridlund}, {Gandolfi}, {Gillon}, {G{\"u}del}, {G{\"u}nther}, {Helling}, {Hoyer}, {Isaak}, {Kiss}, {Kopp}, {Lam}, {Laskar}, {Lecavelier des Etangs}, {Magrin}, {Maxted}, {Mordasini}, {Munari}, {Nascimbeni}, {Olofsson}, {Ottensamer}, {Pall{\'e}}, {Peter}, {Piotto}, {Pollacco}, {Ragazzoni}, {Rando}, {Rauer}, {Reimers}, {Ribas}, {Rieder}, {Santos}, {S{\'e}gransan}, {Smith}, {Stalport}, {Steller}, {Szab{\'o}}, {Thomas}, {Udry}, {Venturini}, \& {Walton}}]{Pagano2024}
{Pagano}, I., {Scandariato}, G., {Singh}, V., {et~al.} 2024, \aap, 682, A102, \dodoi{10.1051/0004-6361/202346705}

\bibitem[{{Palethorpe} {et~al.}(2024){Palethorpe}, {John}, {Mortier}, {Davoult}, {Wilson}, {Rice}, {Cameron}, {Alibert}, \& {Buchhave}}]{Palethorpe2024}
{Palethorpe}, L., {John}, A.~A., {Mortier}, A., {et~al.} 2024, \mnras, 529, 3323, \dodoi{10.1093/mnras/stae707}

\bibitem[{{Parc} {et~al.}(2024){Parc}, {Bouchy}, {Venturini}, {Dorn}, \& {Helled}}]{Parc2024}
{Parc}, L., {Bouchy}, F., {Venturini}, J., {Dorn}, C., \& {Helled}, R. 2024, \aap, 688, A59, \dodoi{10.1051/0004-6361/202449911}

\bibitem[{{Petrovich}(2015)}]{Petrovich2015}
{Petrovich}, C. 2015, \apj, 808, 120, \dodoi{10.1088/0004-637X/808/2/120}

\bibitem[{{Press} {et~al.}(1992){Press}, {Teukolsky}, {Vetterling}, \& {Flannery}}]{Press1992}
{Press}, W.~H., {Teukolsky}, S.~A., {Vetterling}, W.~T., \& {Flannery}, B.~P. 1992, {Numerical recipes in C. The art of scientific computing}

\bibitem[{{Raymond} {et~al.}(2018){Raymond}, {Boulet}, {Izidoro}, {Esteves}, \& {Bitsch}}]{Raymond2018}
{Raymond}, S.~N., {Boulet}, T., {Izidoro}, A., {Esteves}, L., \& {Bitsch}, B. 2018, \mnras, 479, L81, \dodoi{10.1093/mnrasl/sly100}

\bibitem[{{Rein} \& {Liu}(2012)}]{ReinLiu2012}
{Rein}, H., \& {Liu}, S.~F. 2012, \aap, 537, A128, \dodoi{10.1051/0004-6361/201118085}

\bibitem[{{Ricard} \& {Chambat}(2024)}]{Ricard2024}
{Ricard}, Y., \& {Chambat}, F. 2024, \apj, 967, 163, \dodoi{10.3847/1538-4357/ad4113}

\bibitem[{{Rodriguez} {et~al.}(2018){Rodriguez}, {Becker}, {Eastman}, {Hadden}, {Vanderburg}, {Khain}, {Quinn}, {Mayo}, {Dressing}, {Schlieder}, {Ciardi}, {Latham}, {Rappaport}, {Adams}, {Berlind}, {Bieryla}, {Calkins}, {Esquerdo}, {Kristiansen}, {Omohundro}, {Schwengeler}, {Stassun}, \& {Terentev}}]{Rodri2018AJ}
{Rodriguez}, J.~E., {Becker}, J.~C., {Eastman}, J.~D., {et~al.} 2018, \aj, 156, 245, \dodoi{10.3847/1538-3881/aae530}

\bibitem[{{Ros{\'a}rio} {et~al.}(2024){Ros{\'a}rio}, {Demangeon}, {Barros}, {Gandolfi}, {Egger}, {Serrano}, {Osborn}, {Beck}, {Benz}, {Flor{\'e}n}, {Guterman}, {Wilson}, {Alibert}, {Fossati}, {Hooton}, {Delrez}, {Santos}, {Sousa}, {Bonfanti}, {Salmon}, {Adibekyan}, {Nigioni}, {Venturini}, {Alonso}, {Anglada}, {Asquier}, {B{\'a}rczy}, {Barrado Navascues}, {Barrag{\'a}n}, {Baumjohann}, {Beck}, {Billot}, {Biondi}, {Bonfils}, {Borsato}, {Brandeker}, {Broeg}, {Cessa}, {Charnoz}, {Collier Cameron}, {Csizmadia}, {Cubillos}, {Davies}, {Deleuil}, {Deline}, {Demory}, {Ehrenreich}, {Erikson}, {Esposito}, {Fortier}, {Fridlund}, {Gillon}, {G{\"u}del}, {G{\"u}nther}, {Helling}, {Hoyer}, {Isaak}, {Kiss}, {Lam}, {Laskar}, {Lecavelier des Etangs}, {Lendl}, {Luntzer}, {Magrin}, {Maxted}, {Mordasini}, {Nascimbeni}, {Olofsson}, {Osborne}, {Ottensamer}, {Pagano}, {Pall{\'e}}, {Peter}, {Piotto}, {Pollacco}, {Queloz}, {Ragazzoni}, {Rando}, {Rauer}, {Ribas}, {Scandariato}, {S{\'e}gransan}, {Simon}, {Smith}, {Stalport}, {Szab{\'o}},
  {Thomas}, {Udry}, {Van Eylen}, {Van Grootel}, {Villaver}, {Walter}, \& {Walton}}]{Rosario2024}
{Ros{\'a}rio}, N.~M., {Demangeon}, O.~D.~S., {Barros}, S.~C.~C., {et~al.} 2024, \aap, 686, A282, \dodoi{10.1051/0004-6361/202347759}

\bibitem[{{Sanchis-Ojeda} {et~al.}(2014){Sanchis-Ojeda}, {Rappaport}, {Winn}, {Kotson}, {Levine}, \& {El Mellah}}]{Sanchis2014}
{Sanchis-Ojeda}, R., {Rappaport}, S., {Winn}, J.~N., {et~al.} 2014, \apj, 787, 47, \dodoi{10.1088/0004-637X/787/1/47}

\bibitem[{{Shevchenko}(2022)}]{Shevchenko2022}
{Shevchenko}, I.~I. 2022, \mnras, 515, 3996, \dodoi{10.1093/mnras/stac1979}

\bibitem[{{Singh} {et~al.}(2024){Singh}, {Scandariato}, {Smith}, {Cubillos}, {Lendl}, {Billot}, {Fortier}, {Queloz}, {Sousa}, {Csizmadia}, {Brandeker}, {Carone}, {Wilson}, {Akinsanmi}, {Patel}, {Krenn}, {Demangeon}, {Bruno}, {Pagano}, {Hooton}, {Cabrera}, {Santos}, {Alibert}, {Alonso}, {Asquier}, {B{\'a}rczy}, {Navascues}, {Barros}, {Baumjohann}, {Beck}, {Beck}, {Benz}, {Bergomi}, {Bonfanti}, {Bonfils}, {Borsato}, {Broeg}, {Charnoz}, {Cameron}, {Davies}, {Deleuil}, {Deline}, {Delrez}, {Demory}, {Ehrenreich}, {Erikson}, {Fossati}, {Fridlund}, {Gandolfi}, {Gillon}, {G{\"u}del}, {G{\"u}nther}, {Harre}, {Heitzmann}, {Helling}, {Hoyer}, {Isaak}, {Kiss}, {Lam}, {Laskar}, {des Etangs}, {Magrin}, {Maxted}, {Mischler}, {Mordasini}, {Nascimbeni}, {Olofsson}, {Ottensamer}, {Pall{\'e}}, {Peter}, {Piotto}, {Pollacco}, {Ragazzoni}, {Rando}, {Rauer}, {Ribas}, {Salmon}, {S{\'e}gransan}, {Simon}, {Stalport}, {Steinberger}, {Szab{\'o}}, {Thomas}, {Udry}, {Ulmer}, {Van Grootel}, {Venturini}, {Villaver}, {Walton}, \&
  {Zingales}}]{Singh2024}
{Singh}, V., {Scandariato}, G., {Smith}, A.~M.~S., {et~al.} 2024, \aap, 683, A1, \dodoi{10.1051/0004-6361/202347533}

\bibitem[{{Smith} \& {Csizmadia}(2022)}]{Smith2022}
{Smith}, A. M.~S., \& {Csizmadia}, S. 2022, \aj, 164, 21, \dodoi{10.3847/1538-3881/ac704c}

\bibitem[{{Smith} {et~al.}(2017{\natexlab{a}}){Smith}, {Morris}, {Jenkins}, {Bryson}, {Caldwell}, \& {Girouard}}]{smith2017}
{Smith}, J.~C., {Morris}, R.~L., {Jenkins}, J.~M., {et~al.} 2017{\natexlab{a}}, {Kepler Data Processing Handbook: Finding Optimal Apertures in Kepler Data}, Kepler Science Document KSCI-19081-002, id. 7. Edited by Jon M. Jenkins.

\bibitem[{{Smith} {et~al.}(2017{\natexlab{b}}){Smith}, {Stumpe}, {Jenkins}, {Van Cleve}, {Girouard}, {Kolodziejczak}, {McCauliff}, {Morris}, \& {Twicken}}]{smith_pdc2017}
{Smith}, J.~C., {Stumpe}, M.~C., {Jenkins}, J.~M., {et~al.} 2017{\natexlab{b}}, {Kepler Data Processing Handbook: Presearch Data Conditioning}, Kepler Science Document KSCI-19081-002, id. 8. Edited by Jon M. Jenkins.

\bibitem[{{Teixeira} \& {Ballard}(2023)}]{Teixeira2023}
{Teixeira}, K., \& {Ballard}, S. 2023, \apj, 953, 50, \dodoi{10.3847/1538-4357/acdc20}

\bibitem[{{Thomas} \& {Madhusudhan}(2016)}]{Thomas2016}
{Thomas}, S.~W., \& {Madhusudhan}, N. 2016, \mnras, 458, 1330, \dodoi{10.1093/mnras/stw321}

\bibitem[{{Tsiaras} {et~al.}(2016){Tsiaras}, {Waldmann}, {Rocchetto}, {Varley}, {Morello}, {Damiano}, \& {Tinetti}}]{Tsiaras2016ApJ}
{Tsiaras}, A., {Waldmann}, I.~P., {Rocchetto}, M., {et~al.} 2016, \apj, 832, 202, \dodoi{10.3847/0004-637X/832/2/202}

\bibitem[{{Valencia} {et~al.}(2013){Valencia}, {Guillot}, {Parmentier}, \& {Freedman}}]{Valencia2013}
{Valencia}, D., {Guillot}, T., {Parmentier}, V., \& {Freedman}, R.~S. 2013, \apj, 775, 10, \dodoi{10.1088/0004-637X/775/1/10}

\bibitem[{{Veras} {et~al.}(2017){Veras}, {Georgakarakos}, {Dobbs-Dixon}, \& {G{\"a}nsicke}}]{Veras2017}
{Veras}, D., {Georgakarakos}, N., {Dobbs-Dixon}, I., \& {G{\"a}nsicke}, B.~T. 2017, \mnras, 465, 2053, \dodoi{10.1093/mnras/stw2699}

\bibitem[{{Virtanen} {et~al.}(2020){Virtanen}, {Gommers}, {Oliphant}, {Haberland}, {Burovski}, {Weckesser}, {Reddy}, {Cournapeau}, {Nelson}, {alexbrc}, {Roy}, {Peterson}, {Polat}, {Wilson}, {endolith}, {Mayorov}, {van der Walt}, {Brett}, {Laxalde}, {Larson}, {Sakai}, {Millman}, {Colley}, {Lars}, {peterbell10}, {Carey}, {van Mulbregt}, {Bowhay}, {eric-jones}, \& {Striega}}]{Virtanen2020}
{Virtanen}, P., {Gommers}, R., {Oliphant}, T.~E., {et~al.} 2020, Nature Methods, 17, 261, \dodoi{10.1038/s41592-019-0686-2}

\bibitem[{{Vivien} {et~al.}(2024){Vivien}, {Hoyer}, {Deleuil}, {Sulis}, {Santerne}, {Christiansen}, {Hardegree-Ullman}, \& {Lopez}}]{Vivien2024}
{Vivien}, H.~G., {Hoyer}, S., {Deleuil}, M., {et~al.} 2024, \aap, 688, A192, \dodoi{10.1051/0004-6361/202348013}

\bibitem[{{Volk} \& {Malhotra}(2020)}]{Volk2020}
{Volk}, K., \& {Malhotra}, R. 2020, \aj, 160, 98, \dodoi{10.3847/1538-3881/aba0b0}

\bibitem[{{Wang} {et~al.}(2014){Wang}, {Bu}, {Shang}, \& {Gu}}]{Wang2014}
{Wang}, H.-H., {Bu}, D., {Shang}, H., \& {Gu}, P.-G. 2014, \apj, 790, 32, \dodoi{10.1088/0004-637X/790/1/32}

\bibitem[{{Wilson} {et~al.}(2022){Wilson}, {Goffo}, {Alibert}, {Gandolfi}, {Bonfanti}, {Persson}, {Collier Cameron}, {Fridlund}, {Fossati}, {Korth}, {Benz}, {Deline}, {Flor{\'e}n}, {Guterman}, {Adibekyan}, {Hooton}, {Hoyer}, {Leleu}, {Mustill}, {Salmon}, {Sousa}, {Suarez}, {Abe}, {Agabi}, {Alonso}, {Anglada}, {Asquier}, {B{\'a}rczy}, {Barrado Navascues}, {Barros}, {Baumjohann}, {Beck}, {Beck}, {Billot}, {Bonfils}, {Brandeker}, {Broeg}, {Bryant}, {Burleigh}, {Buttu}, {Cabrera}, {Charnoz}, {Ciardi}, {Cloutier}, {Cochran}, {Collins}, {Col{\'o}n}, {Crouzet}, {Csizmadia}, {Davies}, {Deleuil}, {Delrez}, {Demangeon}, {Demory}, {Dragomir}, {Dransfield}, {Ehrenreich}, {Erikson}, {Fortier}, {Gan}, {Gill}, {Gillon}, {Gnilka}, {Grieves}, {Grziwa}, {G{\"u}del}, {Guillot}, {Haldemann}, {Heng}, {Horne}, {Howell}, {Isaak}, {Jenkins}, {Jensen}, {Kiss}, {Lacedelli}, {Lam}, {Laskar}, {Latham}, {Lecavelier des Etangs}, {Lendl}, {Lester}, {Levine}, {Livingston}, {Lovis}, {Luque}, {Magrin}, {Marie-Sainte}, {Maxted}, {Mayo},
  {McLean}, {Mecina}, {M{\'e}karnia}, {Nascimbeni}, {Nielsen}, {Olofsson}, {Osborn}, {Osborne}, {Ottensamer}, {Pagano}, {Pall{\'e}}, {Peter}, {Piotto}, {Pollacco}, {Queloz}, {Ragazzoni}, {Rando}, {Rauer}, {Redfield}, {Ribas}, {Ricker}, {Rieder}, {Santos}, {Scandariato}, {Schmider}, {Schwarz}, {Scott}, {Seager}, {S{\'e}gransan}, {Serrano}, {Simon}, {Smith}, {Steller}, {Stockdale}, {Szab{\'o}}, {Thomas}, {Ting}, {Triaud}, {Udry}, {Van Eylen}, {Van Grootel}, {Vanderspek}, {Viotto}, {Walton}, \& {Winn}}]{Wilson2022}
{Wilson}, T.~G., {Goffo}, E., {Alibert}, Y., {et~al.} 2022, \mnras, 511, 1043, \dodoi{10.1093/mnras/stab3799}

\end{thebibliography}
\bibliographystyle{aasjournal}

\end{document}